\documentclass[twocolumn]{aastex701}
\usepackage{booktabs}
\usepackage{amsmath}
\usepackage{float}
\usepackage{multirow}
\usepackage{longtable}
\usepackage{placeins}
\usepackage{afterpage}
\usepackage{graphicx}
\usepackage{subcaption}
\usepackage{threeparttable}
\usepackage{booktabs}

\newcommand\OVI{O~{\sc vi}}
\newcommand\ovii{O~{\sc vii}}
\newcommand\oviii{O~{\sc viii}}

\newcommand\hi{H {\sc i}}
\newcommand\civ{C {\sc iv}}
\newcommand\kms{$\rm{km\,s^{-1}}$}

\defcitealias{Prochaska2020Joss}{Prochaska et al. 2020, JOSS}
\defcitealias{Prochaska2020Zenodo}{Prochaska et al. 2020, Zenodo}

\shorttitle{GOLIATH Survey-- \OVI\ CGM}
\shortauthors{Flores et al. (2026)}

\begin{document}

\title{The GOLIATH Survey: \OVI\ Absorption Reveals CGM Evolution through the Starburst-to-Quiescent Transition in Massive Galaxies}

\correspondingauthor{Derick Flores, Rongmon Bordoloi}

\author[orcid=0009-0007-3992-6643,gname='Derick',sname='Flores']{Derick Flores}
\affiliation{Department of Physics and Astronomy, North Carolina State University, 2401 Stinson Drive, Raleigh, NC 27695, USA}
\email[show]{daflore3@ncsu.edu}  

\author[orcid=0000-0002-3120-7173,gname='Rongmon', sname='Bordoloi']{Rongmon Bordoloi} 
%%\altaffiliation{}
\affiliation{Department of Physics and Astronomy, North Carolina State University, 2401 Stinson Drive, Raleigh, NC 27695, USA}
\email[show]{rbordol@ncsu.edu}

\author[orcid=0000-0002-7982-412X,gname='Jason', sname='Tumlinson']{Jason Tumlinson} 
%%\altaffiliation{}
\affiliation{Space Telescope Science Institute, 3700 San Martin Drive, Baltimore, MD 21218, USA}
\email{tumlinson@stsci.edu}

\author[orcid=0000-0002-2591-3792,gname='J. Christopher', sname='Howk']{J. Christopher Howk} 
%%\altaffiliation{}
\affiliation{Department of Physics and Astronomy, University of Notre Dame, 225 Nieuwland Science Hall, Notre Dame, IN 46556, USA}
\email{jhowk@nd.edu}

\author[orcid=0000-0001-9158-0829,gname='Nicolas', sname='Lehner']{Nicolas Lehner} 
%%\altaffiliation{}
\affiliation{Department of Physics and Astronomy, University of Notre Dame, 225 Nieuwland Science Hall, Notre Dame, IN 46556, USA}
\email{nlehner@nd.edu}

\author[orcid=0009-0007-9943-1183,gname='Mary', sname='Rickel']{Mary Rickel} 
%%\altaffiliation{}
\affiliation{Department of Physics and Astronomy, University of Notre Dame, 225 Nieuwland Science Hall, Notre Dame, IN 46556, USA}
\email{mrickel@nd.edu}

\author[orcid=0000-0002-3391-2116,gname='Benjamin', sname='Oppenheimer']{Benjamin Oppenheimer}
%%\altaffiliation{}
\affiliation{Center for Astrophysics and Space Astronomy, University of Colorado, 389 UCB, Boulder, CO 80309, USA}
\email{benjamin.oppenheimer@colorado.edu}

\author[orcid=0000-0002-1979-2197,gname='Joseph', sname='Burchett']{Joseph Burchett} 
%%\altaffiliation{}
\affiliation{Department of Physics, New Mexico State University, 1255 N. Horseshoe Drive, Las Cruces, NM 88003, USA}
\email{jnb@nmsu.edu}

\author[orcid=0000-0002-8858-7875,gname='Ahmed', sname='Shaban']{Ahmed Shaban} 
%%\altaffiliation{}
\affiliation{Department of Physics and Astronomy, University of South Carolina, 712 Main Street, Columbia, SC 29208, USA}
\affiliation{Department of Physics and Astronomy, North Carolina State University, 2401 Stinson Drive, Raleigh, NC 27695, USA}
\email{ashaban@sc.edu}

\author[orcid=0000-0002-7893-1054,gname='John', sname='O\'Meara']{John O'Meara} 
%%\altaffiliation{}
\affiliation{W. M. Keck Observatory, 65-1120 Mamalahoa Highway, Kamuela, HI 96743, USA}
\email{jomeara@keck.hawaii.edu}

\author[orcid=0000-0003-0724-4115,gname='Andrew', sname='Fox']{Andrew Fox} 
%%\altaffiliation{}
\affiliation{AURA for ESA, Space Telescope Science Institute, 3700 San Martin Drive, Baltimore, MD 21218, USA}
\email{afox@stsci.edu}

\author[orcid=0000-0002-7738-6875,gname='Jason', sname='Prochaska']{J. Xavier Prochaska}
%%\altaffiliation{}
\affiliation{Department of Astronomy \& Astrophysics, University of California Santa Cruz, 1156 High Street, Santa Cruz, CA 95064, USA}
\email{xavier@ucolick.org}

\author[orcid=0009-0002-1247-0096,gname='Simon Xinlin', sname='Wu']{Simon Xinlin Wu} 
%%\altaffiliation{}
\affiliation{Department of Physics and Astronomy, North Carolina State University, 2401 Stinson Drive, Raleigh, NC 27695, USA}
\email{swu@ncsu.edu}

\author[orcid=0009-0005-8793-2445,gname='Jack', sname='Higginson']{Jack Higginson} 
%%\altaffiliation{}
\affiliation{Department of Physics and Astronomy, North Carolina State University, 2401 Stinson Drive, Raleigh, NC 27695, USA}
\email{}

\author[orcid=0000-0003-3769-9559,gname='Robert', sname='Simcoe']{Robert Simcoe} 
\affiliation{Department of Physics, Massachusetts Institute of Technology, 77 Massachusetts Avenue, Cambridge, MA 02139, USA}
\email{simcoe@space.mit.edu}

\begin{abstract}

We present the GOLIATH survey (Galaxies, Outflows, and the Lifecycle of Immense, Active, Transforming Halos), a study of the multiphase circumgalactic medium (CGM) of massive ($\langle\log M_\star/M_\odot\rangle \approx 11$), blue ($u-r < 1.65$) starburst and post-starburst galaxies at $\langle z\rangle \approx$ 0.43. This work characterizes the warm-hot CGM through \OVI\ absorption in the inner halo ($R/R_{\rm vir} \leq 0.6$) of these rare systems.  Across the star-forming population, \OVI\ column density rises by nearly 1~dex from $\log M_\star/M_{\odot} \sim 8$ to $\sim 11.5$ and increases with specific star-formation rate (sSFR). Two  GOLIATH galaxies with the highest sSFR show the strongest CGM \OVI\ absorption ($\log N_{\rm O\,VI}[\rm cm^{-2}] \gtrsim 15$). In the $\log M_\star/M_{\odot} = [11,12)$ inner-CGM region, massive star-forming galaxies exceed quiescent galaxies on average by a factor of $\sim 3$ in \OVI\ column density and $\sim 1.5$~dex in CGM \OVI\ mass
($\log(M_{\rm O\, VI}/M_\odot) \approx 7.3$ versus $\approx
5.8$), with covering fractions roughly three times higher (62.5\% versus 24.0\% at $\log N_{\rm O\,VI}[\rm cm^{-2}] \geq 14.0$). The \OVI\ line widths and column densities are consistent with feedback-driven radiative cooling, in which outflow shocks heat the CGM and the gas cools back
through the \OVI\ window; the short cooling time, $t_{\rm cool} \sim 10$--$100$~Myr, requires continuous replenishment by active feedback to sustain this reservoir. The residual \OVI\ in quiescent systems may arise from ambient gas at the high-temperature end of the cooling curve. \OVI\ thus traces feedback on short timescales and probes the star-forming--quiescent transition at $\log M_\star/M_{\odot} \gtrsim 11$.

\end{abstract}
%Astro Thesaurus
\keywords{\uat{Galaxies}{573} --- \uat{Circumgalactic Medium}{1879} --- \uat{Extragalactic Astronomy
}{506} --- \uat{Galaxy Evolution}{594} --- \uat{Galaxy Quenching}{2040}}

\section{Introduction}\label{sec:Introduction}

The circumgalactic medium (CGM) is the diffuse, multiphase gas filling a
galaxy's dark matter halo out to roughly the virial radius.  It is the arena in
which galaxies acquire, expel, and recycle their baryons, and it regulates star
formation and quenching over cosmic time. Modern cosmological simulations
reproduce the broad galaxy population well. 
Yet the properties of circumgalactic gas remain the most sensitive to the uncertain subgrid prescriptions for stellar and AGN feedback \citep{Crain_2023,Tumlinson_2017, faucher_2023}. 
This makes the CGM a sensitive probe of how feedback shapes galaxy evolution. Over the past two decades, wide-area galaxy redshift surveys paired with deep absorption-line spectroscopy of background
sources have transformed our understanding of the CGM. 
These studies select galaxy-sightline pairs at
close impact parameters in large numbers. They map the CGM from its cool,
photoionized phase to its warm--hot, collisionally ionized phase, as a function
of galaxy stellar mass, redshift, and star formation rate from local universe to the epoch of reionization \citep{Bordoloi_2011, Steidel__2010, Chen_2010, Tumlinson_2011, Bordoloi_2014, Borthakur_2015, Bordoloi_2018,
Chen_2020, Zahedy_2019_COS_LRG, Wilde_2021, Tchernyshyov_2022, Qu_2024,
Bordoloi_2024, Higginson_2026}.

%\afterpage{\clearpage}
\begin{figure*}[!t]
\centering
\includegraphics[width=\textwidth]{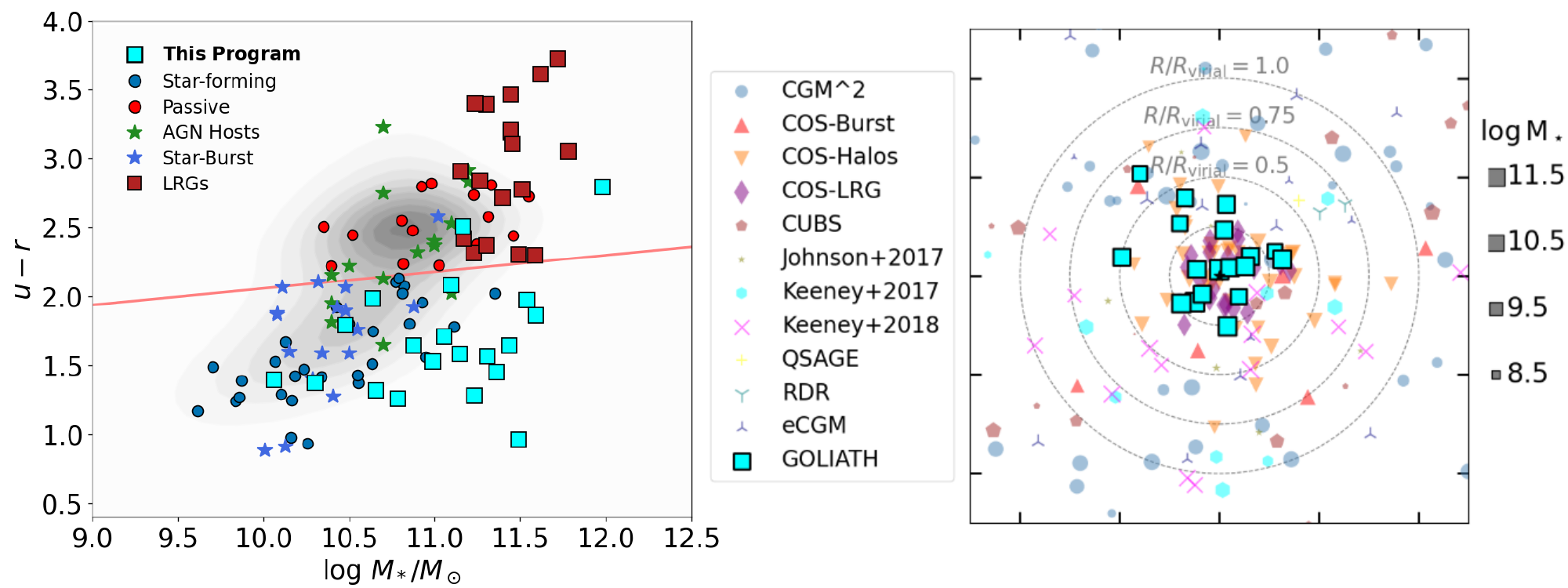}
   \caption{\textit{Left:} Color--mass diagram of GOLIATH survey galaxies compared to archival CGM samples. Background contours show the color--mass distribution of SDSS galaxies from the NASA-Sloan Atlas \citep{Blanton_2011}. The archival star-forming, passive, AGN-host, starburst, and LRG galaxies are drawn from
previous CGM surveys. AGN hosts \citep{Berg_2018} are plotted to mark their location in this plane but are excluded from our analysis, as they lack \OVI\ measurements. \textit{Right:} Radial distribution of the GOLIATH and archival galaxies as a function of $R/R_{\rm vir}$ from the background quasar (central black star). Marker sizes scale with stellar mass. The angles at which the points are plotted are selected randomly.}
    \label{fig:survey_design} 
\end{figure*}

Among UV-accessible ions, \OVI\ is the most sensitive tracer of the warm--hot phase of the CGM. Producing O$^{5+}$ requires $113.9\,\mathrm{eV}$. In collisional ionization equilibrium, its abundance peaks at $T \sim 10^{5.5}\,\mathrm{K}$
\citep{Gnat_Sternberg_2007}. \OVI\ can also arise from a separate channel -- diffuse, low-density gas photoionized by the metagalactic background \citep{Stern_2016, Stern_2019}. 
The \OVI\ doublet (1031, 1037~\AA) is observable only from space at $z \lesssim 2$, beyond which the Lyman-$\alpha$ forest makes it hard to detect.
Around star-forming $\sim L^\star$ galaxies, \OVI\ forms an extensive, high-covering-fraction reservoir. It is detected to $2$--$3$ times the virial radius, declining only shallowly in column density \citep{Tumlinson_2011,
Werk_2016, Tchernyshyov_2022, Qu_2024}. 
This reservoir carries an oxygen mass of $\gtrsim 10^{7}\,M_\odot$, comparable to that in the galaxy itself \citep{Tumlinson_2011}. 
The same temperature phase significantly contributes to the circumgalactic metal budget across a wide range of galaxy mass, down into the dwarf regime \citep{Tchernyshyov_2022,Johnson_2017,
Mishra_2024,Dutta_2025}. 
Around lower-mass dwarfs, \OVI\ remains common and extended, but weakens in column density and covering fraction. This is expected if their
shallow potential wells struggle to retain enriched gas \citep{Johnson_2017,Mishra_2024}. 
The high-ion CGM is instead enhanced around star-bursting hosts.
There, strong \civ\ and  \OVI\ absorption is seen out to $\sim R_{\rm vir}$, with absorber
velocities reaching roughly twice the halo virial velocity. This is direct evidence of
starburst-driven winds \citep{Borthakur_2013, Heckman_2017}. The broad shallow
\OVI\ seen toward some late-type galaxies points further to an extended, hot ($T \sim 10^{6}\,\mathrm{K}$) intragroup medium spanning hundreds of kpc
\citep{Stocke_2013}. In contrast, massive quiescent and luminous red galaxies
show markedly weaker \OVI\ absorption \citep{Tumlinson_2011, Chen_2018, Zahedy_2019_COS_LRG} but a surprisingly large incidence of neutral \hi\ \citep{Thom2012,Bordoloi_2018}.

This dependence of \OVI\ on star formation is observed most dramatically in the
contrast between star-forming and quiescent halos: the \OVI\ bimodality
\citep{Tumlinson_2011}. Its physical origin remains contested and the  models make contradicting predictions. Here we present two scenarios for contrast.  In the ``virial thermometer'' hypothesis
\citep{Oppenheimer_2016}, the deficit around passive galaxies follows from their
greater halo mass. Once the virial temperature exceeds $\sim 10^{6}\,\mathrm{K}$,
\OVI\ is collisionally ionized into higher states (\ovii, \oviii) that radiate
only in the X-ray. Halo mass is therefore the controlling variable, and the model
makes a sharp prediction: star-forming galaxies with
$\log M_\star/M_\odot \gtrsim 11$ should show \OVI\ as weak as their quiescent
counterparts. In the competing ``feedback'' picture, the strong \OVI\ around
star-forming galaxies is instead an excess from recent and ongoing outflows, as
seen directly around starbursts \citep{Heckman_2017}. Here star-formation activity
is the controlling variable, and the prediction inverts: massive star-forming
galaxies should retain \OVI\ as strong as or stronger than in their lower-mass counterparts.

The two scenarios thus diverge only for galaxies that are simultaneously massive and vigorously star-forming. This is precisely the population that existing samples
lack. Within current data, stellar mass and star formation are confounded:
star-forming galaxies are almost exclusively low-mass and quiescent galaxies
high-mass, so both models reproduce the observed bimodality equally well. This
degeneracy is structural, not statistical. Massive galaxies are predominantly
quenched, so high-mass star-forming systems are intrinsically rare, and no survey
has yet assembled a star-forming galaxy sample reaching the stellar masses of luminous red galaxies.
Compounded by the need for a UV-bright quasar at small impact parameter, this
decisive corner of parameter space has remained almost entirely unmeasured.

Here we present the GOLIATH survey (Galaxies, Outflows, and the Lifecycle of Immense,
Active, Transforming Halos) to disentangle these hypotheses. The GOLIATH survey uses {\it HST}/COS FUV spectroscopy of UV-bright quasars at projected distance within $\sim 250\,\mathrm{kpc}$
($\lesssim 0.6\,R_{\rm vir}$) of their hosts. The survey deliberately targets a
rare population: extremely blue, massive galaxies
($ \log M_{\star,~\rm med}/M_\odot \approx 11.1$, $\langle z\rangle \approx 0.43$) caught in
a starburst or post-starburst state. These are likely progenitors of today's
massive red-sequence ellipticals, observed on the eve of quenching. Placing them
on the same \OVI\ -mass--sSFR plane as existing star-forming and quiescent samples
fills the high-mass, star-forming regime where the two scenarios diverge.

This paper, the first in the survey, characterizes the warm--hot CGM of
these massive blue galaxies as traced by \OVI. We present the sample and measure
the \OVI\ absorption strengths along each sightline. We then compare the resulting properties against the star-forming and quiescent loci from the literature (see Table \ref{tab:survey_summary}).
Subsequent papers will extend this characterization to the full ionization and kinematic structure of the gas. 
The selection criteria and survey design are described in Section~\ref{sec:Survey_Desc}; 
our measurement techniques in Section~\ref{sec:Methodology}; 
and our results and discussion of their implications in Sections~\ref{sec:Results} and \ref{sec:Discussion}, respectively. 
Throughout, we assume a flat $\Lambda$CDM cosmology with
$H_0 = 70\,\mathrm{km\,s^{-1}\,Mpc^{-1}}$, $\Omega_\mathrm{M} = 0.3$, and $\Omega_\Lambda=0.7$.

\section{The GOLIATH Survey}\label{sec:Survey_Desc}
We present the GOLIATH survey, designed to characterize the cool and warm-hot CGM of massive galaxies in the final stages of star formation, before and during
the onset of quenching. We target galaxies that are simultaneously massive and show signatures of recent or ongoing star formation. Such systems
occupy the bright, blue corner of the color--mass diagram (CMD; Figure~\ref{fig:survey_design}, left). Galaxies are drawn from the SDSS
photometric catalog, from which we select massive ($\log M_\star/M_\odot > 10.7$), extremely blue ($u-r < 1.65$) systems at $0.2 < z < 0.93$, with a median
stellar mass $ \log M_{\star,~\rm med}/M_\odot \approx 11.1$. In this region of the CMD, galaxies are rapidly star-bursting or in a post-starburst phase with
strong outflows \citep{Tremonti_2007}. Such high-mass star-forming galaxies are the likely progenitors of massive red-sequence galaxies, comparable in stellar
mass and with little growth left before they quench \citep{Peng_2010, Taylor_2015}. This redshift range shifts the \OVI\ $\lambda\lambda 1031, 1037$ doublet and the \hi\ Lyman limit and/or multiple Lyman series lines into the \textit{HST}/COS band-pass. The galaxies were originally selected based on photometric redshifts
and color information from SDSS; a complementary ground-based campaign (Keck/KCWI, LBT/MODS) was subsequently conducted to obtain spectroscopic galaxy redshifts and confirm the sample. This selection strategy, photometric pre-selection followed by spectroscopic confirmation, has been successfully demonstrated by the COS-Halos survey \citep{Tumlinson_2013} and ensures that the sample is assembled without prior knowledge of CGM absorption properties (see section \ref{sec:Methodology}).

For each galaxy, we identify a UV-bright background quasar at small projected separation, drawn from  \citet{Veron_2010} and is cross-matched with GALEX
(FUV $\lesssim 18.9$). The pairs lie within 250 kpc, sampling the inner CGM ($R/R_{\rm vir} \lesssim 0.6$; Figure~\ref{fig:survey_design}, right). The sample comprises 20 quasar sightlines with 12 newly observed and 8 archival that probe the halos of 20 massive blue galaxies. The new spectra were taken with the Cosmic Origins Spectrograph (COS)
on \textit{HST} over 56 orbits (PID 17211, PI: Bordoloi). 
Note that this is a galaxy selected survey, and galaxies are selected without any prior knowledge of absorption in their CGM   \citep[see discussion in][for CGM survey selection philosophy]{Higginson_2026}.

\startlongtable
\begin{deluxetable*}{lcccccccc}
\tabletypesize{\small}
\tablecolumns{9}
\tablewidth{0pt}
\tablecaption{Properties of the GOLIATH survey galaxies.\label{tab:survey_properties}}
\tablehead{
\colhead{QSO ID} & \colhead{$z_{\rm gal}$} & \colhead{$R$} & \colhead{$R/R_{\rm Virial}$} &
\colhead{$\log M_\star$} & \colhead{$\log\,\mathrm{sSFR}$} & \colhead{$u{-}r$} &
\colhead{Grating} & \colhead{Program IDs} \\
\colhead{} & \colhead{} & \colhead{(kpc)} & \colhead{} &
\colhead{($\log M_\odot$)} & \colhead{($\log\,\mathrm{yr}^{-1}$)} &
\colhead{} & \colhead{} & \colhead{}
}
\startdata
  J1319+2728 & 0.6704\tablenotemark{b} & 54.8 & 0.03 & $11.9^{+0.3}_{-0.2}$ & $-12.8<$ & 2.64 & G160M, G185M, E230M & 11667,8672 \\
  J1126+1204 & 0.6503\tablenotemark{a} & 191.8 & 0.18 & $11.6^{+0.2}_{-0.2}$ & $-9.4^{+0.3}_{-0.2}$ & 1.86 & G130M & 13314 \\
  J0956+2515 & 0.3337\tablenotemark{a} & 239.4 & 0.26 & $11.5^{+0.1}_{-0.1}$ & $-16.3<$ & 1.54 & G140L, G130M & 17211 \\
  J1553+3548 & 0.4737\tablenotemark{c} & 155.4 & 0.18 & $11.5^{+0.3}_{-0.2}$ & $-10.6^{+0.8}_{-0.6}$ & 1.28 & G130M, G160M & 11598 \\
  J1145+6206 & 0.5474\tablenotemark{a} & 186.9 & 0.24 & $11.4^{+0.1}_{-0.1}$ & $-9.4^{+0.3}_{-0.2}$ & 1.27 & G140L, G160M & 17211 \\
  J1307+0044 & 0.4677\tablenotemark{a,b} & 97.5 & 0.14 & $11.3^{+0.2}_{-0.1}$ & $-9.8^{+0.2}_{-0.3}$ & 1.49 & G140L, G160M & 17211 \\
  J1212+6153 & 0.4112\tablenotemark{a} & 214.9 & 0.32 & $11.3^{+0.1}_{-0.1}$ & $-9.6^{+0.2}_{-0.2}$ & 1.54 & G140L, G160M & 17211 \\
  J1208+4540 & 0.9274\tablenotemark{a} & 67.3 & 0.12 & $11.2^{+0.3}_{-0.3}$ & $-8.7^{+0.5}_{-0.5}$ & 1.46 & G130M, G160M, & 11741,13846, \\
   &  &  &  &  &  &  & G185M, G225M & 14265,8672 \\
  J0912+2450 & 0.3772\tablenotemark{a} & 68.9 & 0.13 & $11.1^{+0.1}_{-0.1}$ & $-12.2<$ & 2.83 & G140L, G130M & 17211 \\
  J1342-0053 & 0.2271\tablenotemark{a,b} & 35.1 & 0.07 & $11.1^{+0.1}_{-0.1}$ & $-9.8^{+0.1}_{-0.2}$ & 1.66 & G130M, G160M, G185M & 11598,13033,17076 \\
  J1305+5301 & 0.4927\tablenotemark{a,b} & 240.9 & 0.50 & $11.1^{+0.1}_{-0.1}$ & $-9.2^{+0.2}_{-0.3}$ & 2.53 & G140L, G160M & 17211 \\
  J0958+3224 & 0.3982\tablenotemark{a} & 168.4 & 0.36 & $11.0^{+0.1}_{-0.1}$ & $-11.1^{+0.3}_{-0.6}$ & 1.55 & G140L, G160M & 17211 \\
  J0745+1919 & 0.4580\tablenotemark{a} & 97.3 & 0.23 & $11.0^{+0.2}_{-0.1}$ & $-8.8^{+0.2}_{-0.6}$ & 1.51 & G140L, G130M & 17211 \\
  J1244+0755 & 0.2353\tablenotemark{a} & 65.7 & 0.18 & $10.8^{+0.1}_{-0.1}$ & $-9.7^{+0.2}_{-0.2}$ & 1.68 & G140L, G130M & 17211 \\
  J1105+3425 & 0.2895\tablenotemark{a} & 132.8 & 0.43 & $10.8^{+0.1}_{-0.1}$ & $-10.3^{+0.1}_{-0.1}$ & 1.54 & G230L, G130M, G160M & 11541 \\
  J1240+0949 & 0.3093\tablenotemark{a} & 156.3 & 0.61 & $10.7^{+0.1}_{-0.1}$ & $-9.7^{+0.1}_{-0.1}$ & 1.32 & G130M & 11698 \\
  J1405+4704 & 0.3569\tablenotemark{c} & 32.6 & 0.14 & $10.6^{+0.1}_{-0.1}$ & $-9.7^{+0.1}_{-0.1}$ & 1.74 & G140L, G130M, HIRES & 13862,11728 \\
  J0212+0100 & 0.2534\tablenotemark{c} & 68.3 & 0.33 & $10.4^{+0.1}_{-0.1}$ & $-10.3^{+0.1}_{-0.1}$ & 1.81 & G140L, G130M & 17211 \\
  J1407+2933 & 0.2232\tablenotemark{a} & 120.4 & 0.65 & $10.3^{+0.1}_{-0.1}$ & $-9.8^{+0.2}_{-0.3}$ & 1.60 & G140L, G160M & 17211 \\
  J0909+0121 & 0.5360\tablenotemark{c} & 43.5 & 0.30 & $10.0^{+0.2}_{-0.2}$ & $-9.0^{+0.5}_{-0.4}$ & 1.98 & G140L, G160M & 17211 \\
\enddata
\tablenotetext{a}{Spectroscopic Redshift from LBT/MODS}
\tablenotetext{b}{Spectroscopic Redshift from KECK/KCWI/KCRM}
\tablenotetext{c}{Spectroscopic Redshift from SDSS}
\tablecomments{Stellar mass and specific star-formation rate errors are the
  16th/84th percentile bounds from Prospector SED fitting.
%  A machine-readable version of this table is available at [INSERT].
}
\end{deluxetable*}

The COS observations for program 17211 follow a two-part strategy, with two setups per quasar to span the cool and warm-hot phases of the CGM. Each sightline is observed with
the G140L grating, which captures the Lyman limit and Ly$\alpha$ and constrains the \hi\ column density in each galaxy's CGM. Each is also observed with a
medium-resolution ($R \sim 18{,}000$) G130M or G160M grating (chosen by redshift), which resolves \OVI\ $\lambda 1031, 1037$, \hi\ $\lambda 1025$,
C III $\lambda 977$, and C II $\lambda 1036$ at Full  Width Half Maximum (FWHM) $15$--$20\,\mathrm{km\,s^{-1}}$. This yields component structure, kinematics, and accurate \OVI\ columns. We reach $\mathrm{S/N} \sim 10$ at \OVI\ $\lambda 1031$, for an \OVI\ detection threshold of $\log N \sim 13.5$. Several archival sightlines are also covered by NUV observations which allows for high-z \OVI\ detection in one target (Table \ref{tab:survey_properties}). For archival systems, we have access to either the Lyman limit or multiple Lyman series lines in the archival data.

These quasar-galaxy pairs probe the inner CGM, where the interplay between the CGM and the interstellar medium (ISM) is most direct. The selection places our galaxies at substantially smaller median $R/R_{\rm vir}$ than previous surveys of comparably massive systems (Figure~\ref{fig:survey_design}, right; Table \ref{tab:survey_summary}). Crucially, it samples a region of the $M_\star$--SFR plane not previously probed in the CGM: massive galaxies whose only remaining evolutionary step is quiescence. Their stellar masses enable a direct comparison with massive quiescent galaxies
of similar stellar masses, such as the luminous red galaxies (LRGs) \citep[e.g.,][]{Zahedy_2019_COS_LRG}. 
For the remainder of this paper we focus on the warm--hot \OVI\ CGM of these galaxies; subsequent papers will address their cool and hot phases. The galaxy properties and observation details are presented in Table \ref{tab:survey_properties}.

Two caveats should be noted. First, some of these massive galaxies may reside in overdense environments, which we do not address in this paper. This is acceptable for our purpose: we probe the inner CGM, and compare the CGM of these massive galaxies with systems such as LRGs that likely occupy similar overdensities. Second, two of the archival sightlines (J1126+1204 and J1240+0949) lack COS coverage of \OVI\ at the redshift of the target galaxy. We report the galaxy properties of these two galaxies, which will be used for analysis of cool CGM gas in future papers.

\section{Methodology}\label{sec:Methodology}
In this section, we describe the methodology adopted to analyze both the galaxy and the background quasar data, respectively. 

\subsection{Galaxy Redshifts}

Spectroscopic redshifts for each galaxy were obtained by observations using either Keck/KCWI/KCRM (4 galaxies) or LBT/MODS (15 galaxies) instruments. These are marked correspondingly in Table \ref{tab:survey_properties}. Additionally, we use the reported spectroscopic redshifts from SDSS \citep{Abdurro_2022} for J1553$+$3548, J0906$+$015, J1405$+$4704, and J0212$+$0100.

\textit{LBT/MODS}: We obtained longslit optical spectra of 15 galaxies with the Multi-Object Dual Spectrographs (MODS) on the Large Binocular Telescope (LBT) between March 2023 and July 2025 (see \cite{Pogge2010} for a description of the instruments). We simultaneously obtained spectra with MODS1 and MODS2.\footnote{Our observations of J0745$+$1919 used only MODS1 due to technical issues with MODS2.} The MODS instruments are identical double spectrographs providing wavelength coverage from $\lambda \approx 3200$ to $10000$ \AA, with a dichroic splitting the light at $\lambda \sim 5500$ \AA, sampled at 0.5 \AA\ per pixel.\footnote{Our analyses of J1307$+$0044, J1126$+$1205, and J1212$+$6153 make use only of the red-side data, as the blue side did not provide useable data for these observations.} We employed the $5 \times 1\farcs0$ longslit mask, which consists of five consecutive $1\arcmin$ long slits with a few arcsecond gaps between them (though for our purposes, this serves as a traditional longslit spectrograph), and aligned the slits with the parallactic angle. We used the G400L grating (400 lines mm$^{-1}$) on the blue sides and the G670L grating (250 lines mm$^{-1}$) on the red sides. This setup yields a resolution $R \sim 1200$ for both channels. We observed each source with 5 to 7 exposures, integrating 400 to 900 seconds per exposure depending on the target magnitude. 

For reduction and spectral extractions, we use the semi-automated \texttt{Pypeit} reduction software (v1.173 and v1.174; \citetalias{Prochaska2020Joss}, \citetalias{Prochaska2020Zenodo})\footnote{\url{https://pypeit.readthedocs.io/en/stable/}}, using \texttt{Pypeit}'s MODS-specific default configurations. The reduced spectra are flux calibrated using the standard star frames and the \texttt{Pypeit} generated sensitivity function. 

We determine the spectroscopic redshifts of the MODS-targeted galaxies using the redshift-determination code \texttt{REDROCK}\footnote{\url{https://github.com/desihub/redrock}}, created by the Dark Energy Spectroscopic Instrument (DESI; \cite{Aghamousa2016}) collaboration. \texttt{REDROCK} fits a linear combination of Principal Component Analysis-based spectral templates and identifies the best fit as the redshift and template that minimizes $\chi^{2}$ (see \cite{Ross2020} for more details on \texttt{REDROCK}’s functionality). We use only the galaxy spectral templates for the fits. For the MODS data, the most important spectral features are typically the H$\beta$ and [\ion{O}{3}] emission lines, the [\ion{O}{2}] lines near 3727 \AA, and prominent stellar absorption features such as \ion{Ca}{2} and the Balmer convergence. We report for the 15 GOLIATH targets the best fit redshifts in Table \ref{tab:survey_properties}. 

\textit{KECK}: We obtain integral field unit (IFU) observations for 4 galaxies (J1305$+$5301, J1307$+$0044, J1319$+$2728, and J1342$-$0053) using Keck Cosmic Web Imager \citep[KCWI;][]{morrissey_2018keck} and Keck Cosmic Reionization Mapper \citep[KCRM;][]{McGurk_2024KCRM}. The KCWI and KCRM observations were taken simultaneously on the night of April 4, 2024 (program ID: N110, PI: Bordoloi), using the small slicer with the large BL/RL gratings for the blue and red channels, respectively. Each target has a 1390s exposure for the blue KCWI, and 4 exposures from the red KCRM with a total integration time of 1200s ($4\times 300$s). Additional follow-up observations of the target J1319$+$2728 were obtained with KCWI/KCRM using the Small BL/RL setup on the night of December 14, 2025 (program ID: U065, PI: Treu). The total exposure time for this target on that night is 2060s ($2\times 1030$s) for the blue KCWI, and 1800s ($6\times 300$s) for the red KCRM. This setup has a field of view of $8.4^{\prime\prime} \times 20.4^{\prime\prime}$, slice width of $0.35^{\prime\prime}$, and spectral resolutions of $R\sim 3600$ for the blue channel KCWI and $R\sim 2000$ for the red channel KCRM. This setup has a wavelength coverage of $\lambda \sim 3690 -5690${\AA} for KCWI for both nights. For KCRM, the wavelength coverage is $\lambda \sim 5480 - 8750${\AA} for the 2024 observations, and $\lambda \sim 5650 -8950${\AA} for the 2025 observation. This is due to the choice of slightly different central wavelengths for KCRM for each night. For the blue KCWI data, we use the official KCWI data reduction pipeline \texttt{KCWI\_DRP}\footnote{https://kcwi-drp.readthedocs.io/en/latest/} v1.2.0 to produce final flux-calibrated datacubes for individual exposures using observations of standard flux calibration stars taken during the same night of obervation. For the red KCRM data, we use the software package \texttt{kcwikit}\footnote{https://github.com/yuguangchen1/KcwiKit} \citep{Chen2021KBSSKCWI, Prusinski2024kcwikit}, which is a modified version of the official pipeline and provides better control on cosmic rays removal and sky-subtraction in the red cubes. In addition, we use \texttt{kskywizard}\footnote{https://github.com/zhuyunz/KSkyWizard} to perform telluric correction and remove residual sky contribution using Zurich Atmosphere Purge algorithm \citep[ZAP;][]{Soto_2016ZAP}. We use \texttt{Montage} \citep{Jacob_2010montage, Jacob2010montagecode} to reproject the exposures into a grid of square pixels and coadd individual exposures into final co-added datacubes.

For each galaxy datacube, we extract a 1D spectrum from an aperture in the co-added cube that captures the entire light of each galaxy. Redshift was estimated by fitting a double Gaussian to the [\ion{O}{2}] emission line doublet. We do not use \texttt{REDROCK} as with the MODS observations, as we only have access to this single line complex.

Three of our galaxies were observed by both MODS and KCWI. For these systems, we utilize the redshift obtained with MODS since it contains more emission lines. The KCWI data for these three targets is used as a comparison and to check for consistency.

\subsection{Galaxy Stellar Masses and Star-Formation Rates}
After measuring spectroscopic redshifts we estimate stellar masses and star-formation rates from spectral energy distribution (SED) fitting to the broad band photometry of the galaxies. 

We fit these data with the \texttt{Prospector} SED-fitting package \citep{Prospector_2021, Leja_2017}, following a procedure similar to \citet{Matthee_2023} and \citet{Shaban_2025},
with unconstrained emission-line photometry. The free parameters are stellar mass, stellar and gas-phase metallicity, galaxy age, star-formation history, dust
attenuation, and the ionization parameter. We fix each galaxy's redshift and the SED fit is performed at that redshift.

We take the stellar mass and SFR from the resulting posteriors, with uncertainties from the 16$\rm{^{th}}$ and 84$\rm{^{th}}$ percentiles. We convert stellar mass to halo
mass using the \textsc{UniverseMachine} stellar-mass--halo-mass relation \citep{Behroozi_2019}. We adopt $R_{200c}$ as the virial radius, the radius within which the mean density is 200 times the critical density
of the Universe \citep[following methodology outlined in][]{Bordoloi_2024}. In the rest of the paper we refer to $R_{200c}$ as the $\rm{R_{vir}}$ for the galaxy. For simplicity, we take a binary cut on galaxy's specific star-formation rate (sSFR)  and classify a galaxy as
star-forming if its satisfies $\mathrm{sSFR} \geq 10^{-11}\,\mathrm{yr^{-1}}$, and quiescent otherwise. These properties are summarized in Table~\ref{tab:survey_properties}.

\begin{deluxetable*}{lcccc}
\tablecaption{Summary properties of the comparison surveys included in this work.\label{tab:survey_summary}}
\tablehead{
\colhead{Survey} & \colhead{$N_{\rm gal}$} & \colhead{Median $R/R_{\rm vir}$} &
  \colhead{Median $\log M_\star$} & \colhead{Reference} \\
\colhead{} & \colhead{} & \colhead{} & \colhead{($\log M_\odot$)} & \colhead{}
}
\startdata
  GOLIATH & 18 & 0.24 & 11.1 & This Work \\
  CGM$^2$ & 126 & 1.84 & 9.2 & \cite{Tchernyshyov_2022} \\
  COS-Burst & 6 & 0.68 & 10.5 & \cite{Heckman_2017} \\
  COS-Halos & 39 & 0.32 & 10.5 & \cite{Werk_2016} \\
  COS-LRG & 19 & 0.17 & 11.2 & \cite{Zahedy_2019_COS_LRG} \\
  CUBS & 50 & 1.72 & 9.5 & \cite{Qu_2024} \\
  Johnson+2017 & 11 & 0.80 & 8.2 & \cite{Johnson_2017}\tablenotemark{a} \\
  Keeney+2017 & 13 & 0.92 & 10.3 & \cite{Keeney_2017}\tablenotemark{a} \\
  Keeney+2018 & 18 & 0.82 & 10.4 & \cite{Keeney_2018}\tablenotemark{a} \\
  QSAGE & 2 & 1.15 & 9.2 & \cite{Bielby_QSAGE_2019}\tablenotemark{a} \\
  RDR & 2 & 0.66 & 11.0 & \cite{Berg_2018}\tablenotemark{a} \\
  eCGM & 33 & 1.46 & 9.6 & \cite{Johnson_eCGM_2015}\tablenotemark{a} \\
  MUSEQuBES & 60 & 1.11 & 8.8 & \cite{Dutta_2025}\tablenotemark{b} \\
\enddata
\tablenotetext{a}{Adopted from the CGM$^2$ data compilation.}
\tablenotetext{b}{MUSEQuBES does not report the individual component column density and is only used for the kinematic analysis.}
\end{deluxetable*}

\subsection{HST/COS Data Reduction}

We retrieve the calibrated one-dimensional spectra (\texttt{x1d} files), processed with \texttt{CALCOS}, from MAST. We inspect these and discard any failed or erroneous exposures. The remaining exposures are coadded with the \texttt{HASP} software \citep{Debes_et_al.}, which combines the \texttt{x1d} files separately
by visit and instrument. We then coadd the per-visit spectra across all visits using inverse-variance weighting, separately for each instrument. The final spectrum is resampled to three pixels per resolution element (Nyquist sampling), maximizing the signal-to-noise ratio while preserving the full kinematic information. These co-added spectra are our final data products for the absorption-line analysis. They reach a typical signal-to-noise ratio of
$10$--$20$ for G130M/G160M and $5$--$11$ for G140L.

\subsection{Absorption Line Analysis}

We perform absorption-line analysis with the \texttt{rbcodes} Python API
\citep{bordoloi_2025_15723701}, which has been effectively applied to spectral data in previous work \citep{Bordoloi_2024, Higginson_2026}. We use its
tools for absorption-line identification, continuum normalization, and equivalent
width ($W_r$) measurements.

We start by inspecting each quasar spectrum to identify absorption from the target galaxy and from intervening systems that could blend with the lines of interest, including milky way absorption lines and lines at the QSO redshift. Because the G140L and G130M/G160M spectra are available together, we identify absorbers in both simultaneously. 

For each sightline, we use the measured spectroscopic galaxy redshift to associate absorption with the target galaxy. We shift the spectrum to the galaxy rest frame and extract a slice within $\pm 1500\,\mathrm{km\,s^{-1}}$ of each ion's rest wavelength. We normalize the continuum with polynomial fits of degree $n = 1$--$7$, selecting the optimal degree via the Bayesian Information Criterion (BIC); for
lower-S/N spectra we use a spline fit instead. All absorption lines are masked during the continuum fit.

From the normalized spectra we measure the equivalent width ($W_r$) of the \OVI\ $\lambda\lambda 1031.9, 1037.6$ doublet, setting the equivalent width integration window with visual inspection. We classify a detection as $W_r > 3\,\sigma_{W_r}$. Where no significant \OVI\ absorption is present within $\pm 1500\,\mathrm{km\,s^{-1}}$, we report a
$3\sigma$ upper limit measured over a $200\,\mathrm{km\,s^{-1}}$ window. One sightline, J1319$+$2728, is treated as an exception; we discuss it in
Section~\ref{sec:Results}. In this work, we use equivalent width measurements to classify detections and non-detections. The $3\sigma$ upper limits are used to compute apparent optical depth (AOD) column density limits for non-detections.

\begin{figure*}[ht]
    \centering
    \includegraphics[width=1\linewidth]{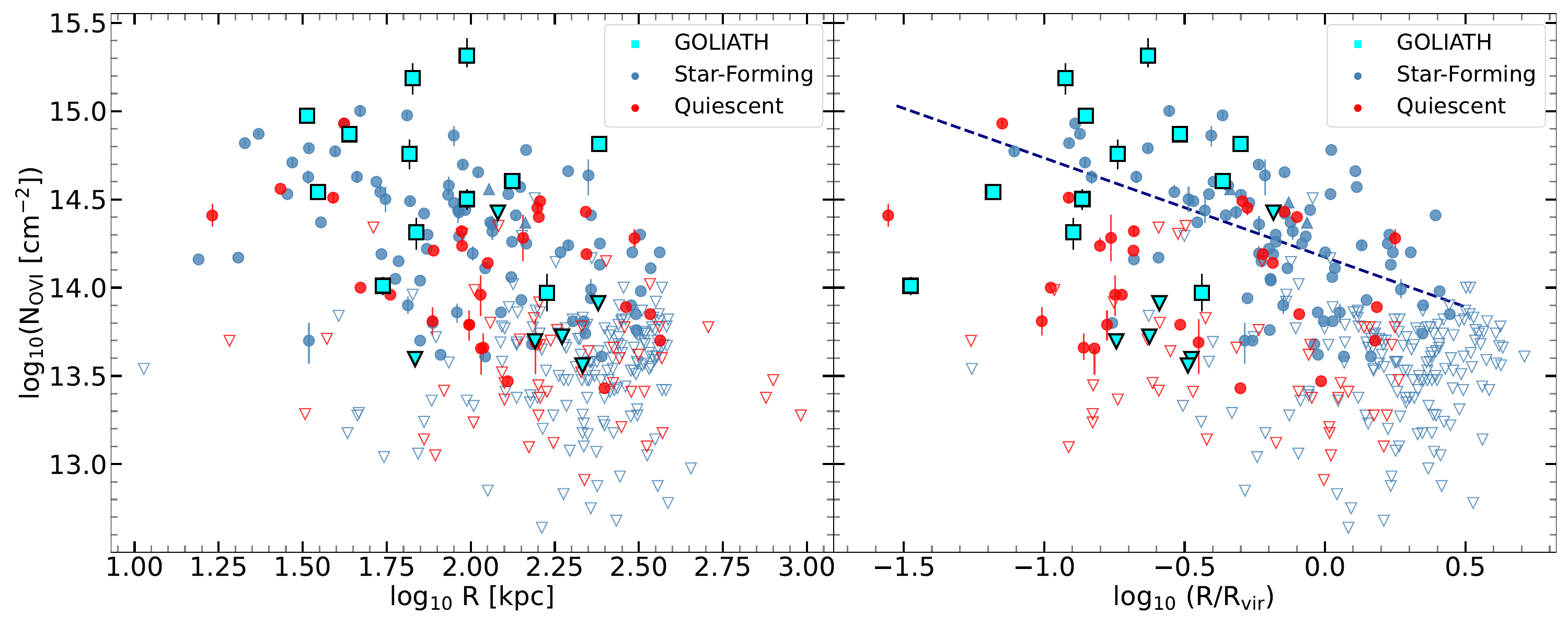}
    \caption{\OVI\ column density as a function of projected distance to the host galaxy. \textit{Left:} \OVI\ column density versus physical projected separation. GOLIATH galaxies (cyan squares) reach \OVI\ columns comparable to star-forming $L^\star$ galaxies (blue circles) but with larger scatter, and lie systematically above passive galaxies (red circles). Non-detections are shown as downward triangles at the $3\sigma$ upper limit. \textit{Right:} the same quantity versus projected separation normalized by the virial radius, $R/R_{\rm vir}$. Within $R/R_{\rm vir} \leq 0.6$, the GOLIATH galaxies show a markedly larger spread in \OVI\ than the star-forming sample.}
    \label{fig:ovi_radial_profile}
\end{figure*}

\subsection{Column Density Measurement}
Column densities are derived with the \texttt{rbvfit} package
\citep{bordoloi_2025_rbvfit}, which fits Voigt profiles to the normalized spectra. We fit the \OVI\ doublet simultaneously across all available instruments, tightening the constraints and yielding more precise column densities. The higher-resolution G130M or G160M spectra set the number of velocity components in the \OVI\ profile, and these spectra are fit jointly with the lower-resolution G140L data. This makes the fits robust: the joint model exploits the full available SNR while retaining the kinematic structure resolved by the medium-resolution spectra \citep[see][for a detailed example of such joint fitting
fits]{Higginson_2026}. Identified blended or intervening absorption falling in the fitted window is modeled jointly with the target lines. The package \texttt{rbvfit}
supports the \texttt{Zeus} MCMC sampler; we use \texttt{Zeus} \citep{karamanis2020ensemble, karamanis2021zeus} throughout this work. Any unidentified
intervening absorption in the spectral slice is treated as a nuisance component and modeled as \hi. These nuisance components absorb the otherwise-unmodeled flux and do not bias the inferred \OVI\ column densities. For each sightline, we compute the total \OVI\ column density by summing the column densities of individual Voigt profile
components. Unless stated otherwise, throughout this paper
we report this integrated \OVI\ column density for each
system. The fitted column densities and Voigt-profile
parameters are listed in Table~\ref{tab:ovi_voigt_bms};
the spectra and profile fits are shown in
Appendix~\ref{fig:ovi_highres}.

\section{Results}\label{sec:Results}

In this section, we present the \OVI\ absorption properties in the CGM of high-mass galaxies from the GOLIATH survey. 

\subsection{\OVI\ Column Density Profile}
We first present the \OVI\ column density variation around the GOLIATH galaxies and compare it with that found around 
other star-forming and passive galaxies. Figure~\ref{fig:ovi_radial_profile} 
shows the \OVI\ column density as a function of impact parameter 
(left panel) and impact parameter normalized to the virial radius 
of the host galaxy ($R/R_{\rm vir}$, right panel). The archival 
galaxy sample is drawn from a list of publications presented in  Table~\ref{tab:survey_summary}. As a function 
of impact parameter, the \OVI\ absorption around GOLIATH galaxies 
(cyan squares) follows a similar trend to archival star-forming 
(blue circles) galaxies and shows elevated absorption strength 
relative to passive (red circles) galaxies. Note that while impact 
parameter reaches 250~kpc, owing to their higher stellar mass, all 
GOLIATH sightlines probe the inner CGM ($R/R_{\rm vir} \leq 0.6$).

\OVI\ absorption strength declines for all galaxies at larger 
$R/R_{\rm vir}$ (Figure~\ref{fig:ovi_radial_profile}, right panel). 
We characterize this column density falloff by fitting a simple power law 
($\gamma=-0.6\pm 0.1$) to the \OVI\ detections 
around star-forming galaxies. The GOLIATH detections scatter about 
this relation with $\sigma = 0.48$~dex, compared to $\sigma = 
0.30$~dex for the literature star-forming sample which is 58\% smaller. 
This excess scatter is most pronounced at $R/R_{\rm vir} \leq 0.6$, 
where GOLIATH sightlines span column densities ranging from values 
typical of star-forming galaxies down to those of quiescent systems. 
We return to the physical interpretation of this scatter in 
Section~\ref{sec:Discussion}.

\begin{figure*}[!t]
    \centering
    \includegraphics[width=0.75\linewidth]{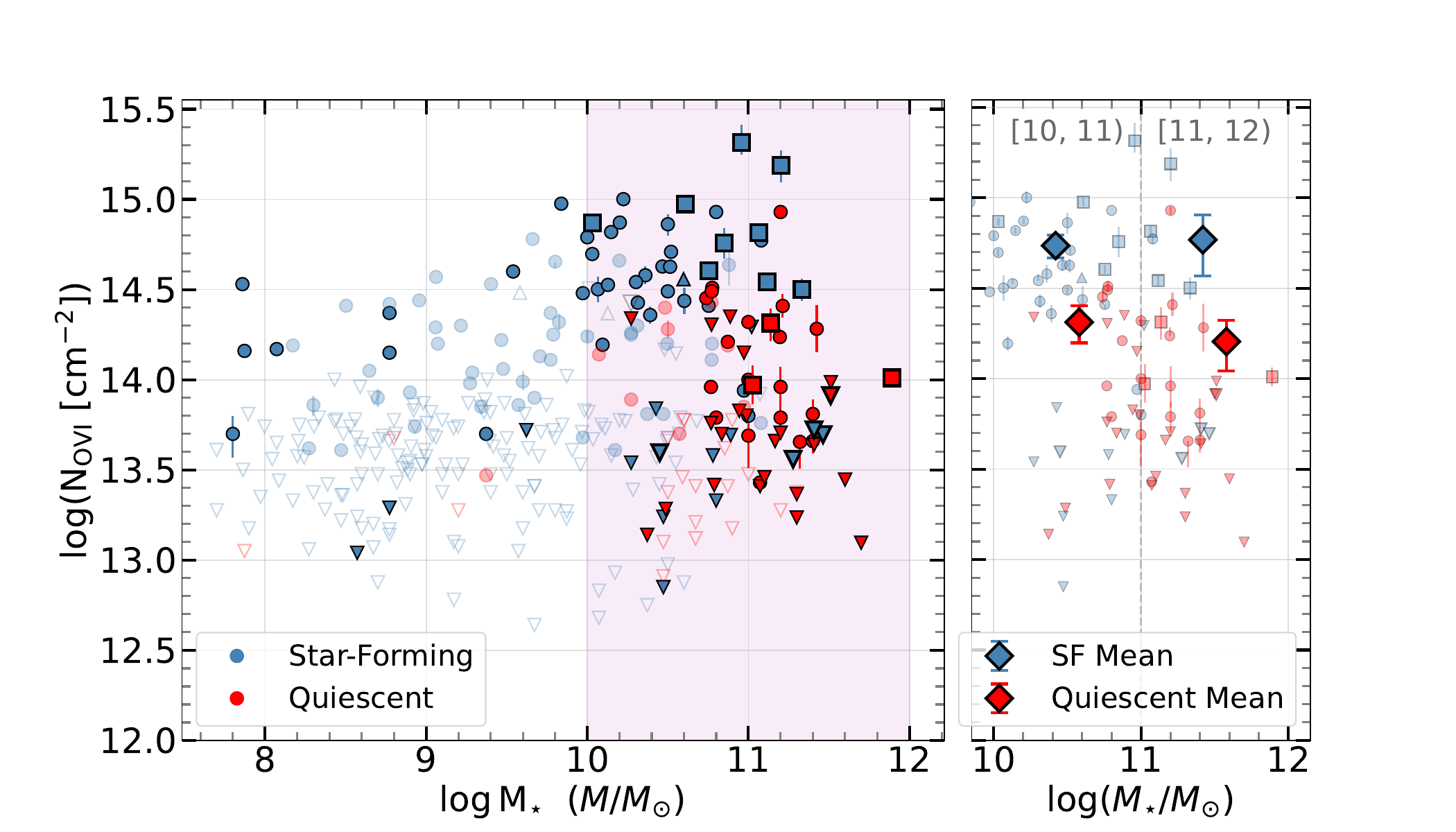}
    \caption{\OVI\ column density as a function of stellar mass for
star-forming (blue) and quiescent (red) galaxies. GOLIATH galaxies
are shown as squares; archival galaxies are shown as circles.
Bolder points with outlines are inner-CGM observations
($R/R_{\rm vir} \leq 0.6$); fainter points lie at larger
$R/R_{\rm vir}$. Upper limits are shown as downward-pointing triangles. The shaded region indicates the mass range used in the right panel for calculating the mean. 
\textit{Right:} Inner-CGM sightlines only, with the mean \OVI\
column density computed for star-forming and quiescent galaxies
in the $\log M_\star/M_{\odot} = [10,11)$ and $[11,12)$ bins. Across both mass bins, inner-CGM
star-forming galaxies have a mean \OVI\ column density
$\sim$0.5~dex higher than quiescent galaxies at the same
stellar mass.}
    \label{fig:logM_logN}
\end{figure*}

\subsection{\OVI\ Absorption Strength With Galaxy Mass}\label{sec:mass_dep}

Figure~\ref{fig:logM_logN} shows \OVI\ column density as a function
of stellar mass for the full combined sample. Across the star-forming
population, $\log N_{\rm O\,VI}$ increases steadily from
$\log M_\star/M_{\odot} \sim 8$ to $\sim 11.5$, rising by nearly 1~dex
independent of whether the full sample or only inner-CGM sightlines
are considered. Quiescent galaxies are systematically offset to lower
column densities at all stellar masses, with a higher fraction of
non-detections.  
A Generalized Kendall's Tau correlation \citep{astrostats} test was conducted to evaluate the null hypothesis that there is no monotonic relationship between the column density ($\log \rm N_{\text O VI}$) and the stellar mass ($\log \rm M_*$). Based on the test, we reject the null hypothesis ($\tau=0.191, p<0.0001$). The results indicate a statistically significant, positive monotonic relationship between both variables. This implies that higher stellar masses are preferentially associated with higher \OVI \, column densities though the relatively low coefficient suggests a weak correlation with substantial scatter in the distribution.

To quantify this offset in the mass regime probed by GOLIATH, we
restrict to inner-CGM sightlines ($R/R_{\rm vir} \leq 0.6$) and
compute the mean \OVI\ column density in two bins
(Figure~\ref{fig:logM_logN}, right). In both the $\log M_\star/M_{\odot} =
[10,11)$ and $[11,12)$ bins, star-forming galaxies show a mean
column density $\sim 0.5$~dex higher than quiescent galaxies of
the same stellar mass.

Most of the GOLIATH galaxies (squares) occupy the high-mass bin
at $\log M_\star/M_{\odot} = [11,12)$. While their mean \OVI\ column density
is consistent with the broader star-forming (SF) population, they exhibit a large
scatter, with individual sightlines spanning from column densities
typical of star-forming galaxies down to values approaching those
of quiescent systems. This spread suggests that the warm CGM in
this mass regime is in an intermediate state of depletion as these
galaxies approach quiescence; we discuss the physical processes
driving this transition in Section~\ref{sec:Discussion}.

\subsection{\OVI\ Absorption Strength With Galaxy sSFR}\label{sec:ssfr_dep}
%\subsection{sSFR Dependence}\label{sec:ssfr_dep}

\begin{figure}[!tb]
    \centering
    \includegraphics[width=1.0\linewidth]{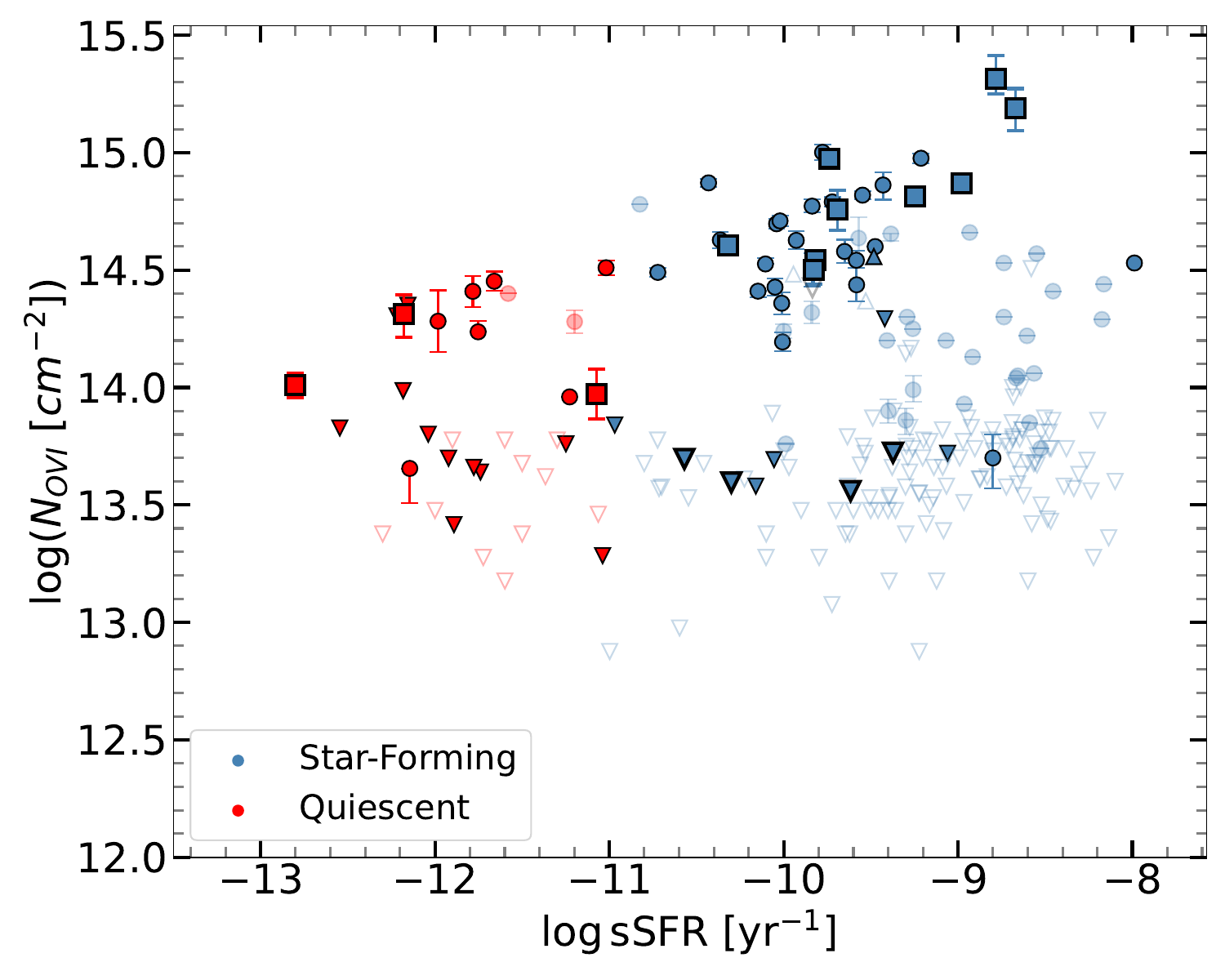}
    \caption{\OVI\ column density as a function of specific star-formation rate (sSFR) for star-forming (blue) and quiescent (red) galaxies. GOLIATH galaxies are shown as squares; archival galaxies are shown as circles. Solid points with outlines are inner-CGM observations ($R/R_{\rm vir} \leq 0.6$); transparent points lie at larger $R/R_{\rm vir}$. Upper limits are shown as downward-pointing triangles. \OVI\ column density increases with sSFR across the combined sample; GOLIATH galaxies occupy the upper locus of this relation, with several sightlines reaching the highest \OVI\ column densities observed in any CGM survey to date.}
    \label{fig:ssfr_logN}
\end{figure}

Figure~\ref{fig:ssfr_logN} shows \OVI\ column density as a function
of sSFR for the combined sample. \OVI\ column density increases with
sSFR across the full population, with quiescent galaxies ($\log\,
\rm sSFR\; [yr^{-1}] \lesssim -11$) showing systematically lower column densities
and a higher fraction of non-detections than star-forming galaxies
at all sSFR. This bimodality, first reported by \citet{Tumlinson_2011},
is clearly recovered in the combined sample. 
We once again conduct a generalized Kendall's Tau correlation test for the column density ($\log \rm N_{\text O VI}$) and the sSFR. We reject the null hypothesis that there is no correlation between sSFR and column density. The test results indicate a small monotonic decrease at small significance ($\tau = -0.068,\,  p < 0.05 $).  This implies that throughout the entire sample, the column density remains similar. The low coefficient and high p-value indicate that the large spread in the entire sSFR distribution leads to a weak correlation. The galaxies observed with a large $R/R_{\rm vir}$ ratio skew the test to have a negative correlation. To remove this bias, we perform the same test on the set of observations with $R/R_{\rm vir}~\leq~0.6$. In this case, we find a strong and statistically significant correlation ($\tau=0.393, ~p<0.0001$) between the column density and sSFR.

Additionally, we perform a Kolmogorov-Smirnov \citep{KStest} test on the inner-CGM quiescent and star-forming samples. We are able to reject the null hypothesis that both samples originate from the same parent distribution ($D=0.58, p<0.001$).
These results directly suggest that the sSFR dichotomy is traceable by the \OVI ~column density.

We then compare the high-mass ($\log M_* >10.7$) star-forming galaxies to lower-mass star-forming galaxies using a log-rank test \citep{logrank}. We find that high-mass galaxies consistently have higher column densities at matched sSFR values($\chi^2=20.6, ~p <0.0001$).

The GOLIATH galaxies (squares) occupy the high-sSFR end of the
star-forming sequence ($\log\,\rm sSFR \;[yr^{-1}]\gtrsim -10$) and span a wide range in \OVI\ column density. Several GOLIATH sightlines reach $\log N_{\rm O\,VI} \gtrsim 15.0$, among the highest column densities observed for any CGM \OVI\ absorber. This is consistent with the high-mass SF bin result from Section~\ref{sec:mass_dep}: GOLIATH galaxies sit at the upper locus of the \OVI--sSFR relation but with substantial scatter, reflecting the same spread seen as a function of stellar mass.  

Only one of the GOLIATH galaxies falls in the green valley
($\log\,\rm sSFR \,[yr^{-1}] \approx -11$; \citealt{Wetzel_2012}),
consistent with their selection as actively star-forming or
post-starburst systems. To quantify the \OVI\ detection rate across
the SF--quiescent divide, we compute the covering fraction in the
following section.

\subsection{\OVI\ Covering Fraction}
\begin{figure}[!ht]
    \centering
    \includegraphics[width=1\linewidth]{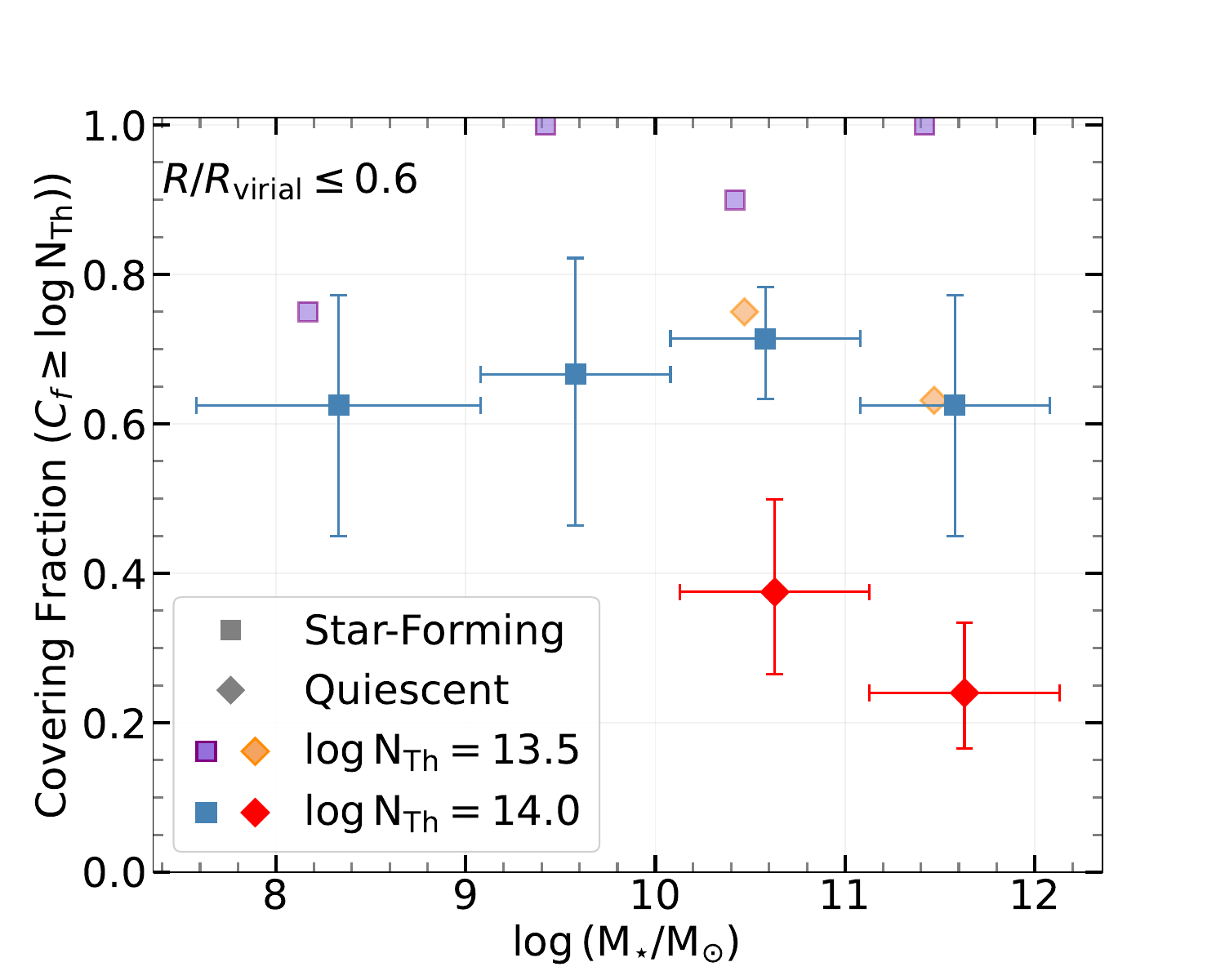}
    \caption{Covering fraction of \OVI\ as a function of stellar mass for inner-CGM sightlines ($R/R_{\rm vir} \leq 0.6$). Star-forming and quiescent galaxies are shown as squares and diamonds, respectively, at two column density thresholds: $\log N_{\rm Th}= 13.5$ (purple/orange), and $\log N_{\rm Th}
= 14.0$ (blue/red). Horizontal range bars
indicate the bin range; vertical error bars are Wilson score
confidence intervals. A small horizontal offset is applied between
thresholds for clarity. At $\log M_\star/M_{\odot} = [11,12)$, star-forming galaxies maintain a covering fraction nearly three times higher than quiescent galaxies of the same mass ($0.63$ versus $0.24$ at $\log N_{\rm Th} = 14.0$).}
    \label{fig:ovi_covering_frac}
\end{figure}

Figure~\ref{fig:ovi_covering_frac} shows the \OVI\ covering fraction
as a function of stellar mass for inner-CGM sightlines
($R/R_{\rm vir} \leq 0.6$), computed at two column density
thresholds. The first, $\log N_{\rm Th} = 13.5$, matches previous
literature \citep{Tumlinson_2011, Tchernyshyov_2022} and enables
a direct comparison. The second, $\log N_{\rm Th} = 14.0$, better
represents the \OVI\ column densities observed in the halos of
massive galaxies, where all GOLIATH $>3\sigma$ detections reside.

At both thresholds, star-forming galaxies maintain higher covering
fractions than quiescent galaxies across all stellar mass bins,
consistent with the column density trends established in
Sections~\ref{sec:mass_dep} and \ref{sec:ssfr_dep}. The contrast
is sharpest at high mass: in the $\log M_\star/M_{\odot} = [11,12)$ bin,
star-forming galaxies have a covering fraction of 62.5$\pm$14\% at
$\log N_{\rm Th}\,[\rm cm^{-2}] = 14.0$, compared to 24$\pm$10\% for quiescent
galaxies which is nearly a factor of three higher. The GOLIATH sample
alone has a covering fraction of 64.7$\pm$11\% at this threshold,
consistent with the broader high-mass SF population.
We perform Fisher's exact test \citep{Fisher_1922} between star-forming and quiescent galaxies with stellar mass $\log M_* > 10.5$. We are able to reject the null hypothesis that the detection rates between the star forming and quiescent samples are the same (odds ratio=4.5, $p<0.01$). This means that above the threshold of $\log N_{th} = 14.0$, star-forming galaxies in this mass range are 4.5 times more likely to show a detection than passive galaxies. 
This test shows that the \OVI\ gas tracks quiescence.

\subsection{\OVI\ Gas Kinematics}\label{sec:kinematics}

\begin{figure*}[!htb]
    \centering
    \includegraphics[width=1.\linewidth]{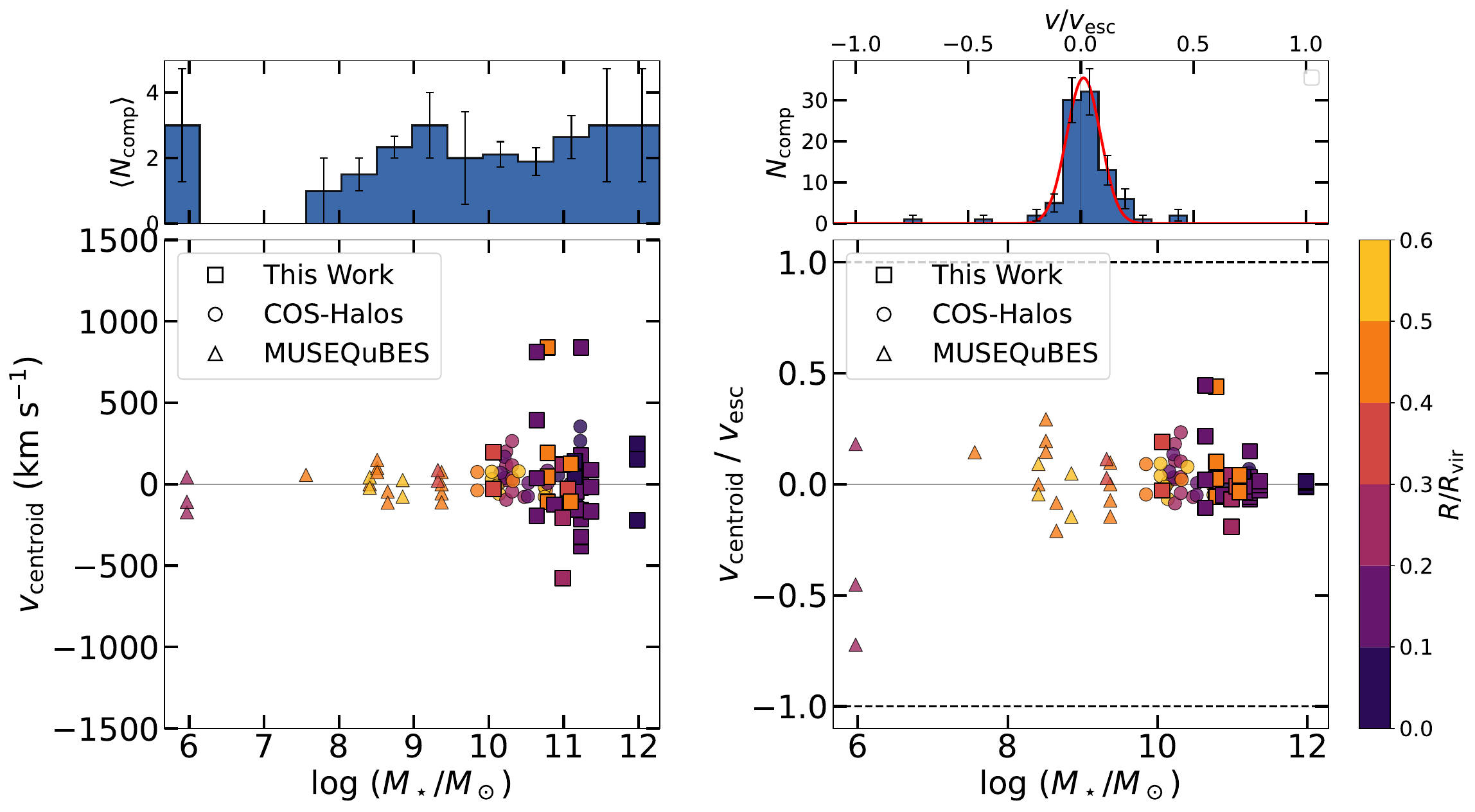}
    \caption{Kinematic properties of \OVI\ absorbers for GOLIATH
(squares), COS-Halos (circles), and MUSEQuBES (triangles)
detections with $R/R_{\rm vir} \leq 0.6$.
\textit{Upper left:} Mean number of \OVI\ kinematic components
$\langle N_{\rm comp}\rangle$ per stellar mass bin for $>3\sigma$
detections. 
\textit{Lower left:} Component velocity centroids as a function
of stellar mass, colored by $R/R_{\rm vir}$.
\textit{Upper right:} Distribution of all component velocities
normalized to the host galaxy escape velocity $v/v_{\rm esc}$,
with a Gaussian fit (red) centered at $0.01\,v_{\rm esc}$ with
$\sigma = 0.08\,v_{\rm esc}$.
\textit{Lower right:} Component velocity centroids normalized
by escape velocity as a function of stellar mass, colored by
$R/R_{\rm vir}$. Dashed lines mark $v/v_{\rm esc} = \pm 1$.
All detected \OVI\ components lie within the escape velocity
of their host galaxy.}
    \label{fig:velocity_plots}
\end{figure*}

Figure~\ref{fig:velocity_plots} shows the kinematic properties of
\OVI\ absorbers for the GOLIATH sample compared to COS-Halos
\citep{Tumlinson_2011} and MUSEQuBES
\citep{Dutta_2025} within
$R/R_{\rm vir} \leq 0.6$.

The velocity centroids of individual components span a wider
range at higher stellar mass (Figure~\ref{fig:velocity_plots},
lower left), reaching a range of $1220$ \kms\ for
GOLIATH galaxies compared to $343$ \kms\
for the COS-Halos and MUSEQuBES samples. Despite this larger velocity spread,
all detected \OVI\ components remain gravitationally bound to
their host galaxies. The distribution of $v/
v_{\rm esc}$ across the full sample (Figure~\ref{fig:velocity_plots},
upper right) is well described by a Gaussian centered at
$0.01\,v_{\rm esc}$ with $\sigma = 0.08\,v_{\rm esc}$,
confirming that the bulk of the warm gas is not escaping its
host halo. No \OVI\ components exceed $|v/v_{\rm esc}| = 1$
in the GOLIATH sample; the few components approaching this
limit in the combined sample are drawn from
\citet{Dutta_2025}. Note that the mean number of \OVI\ kinematic components per sightline appear to increases with stellar mass across the combined sample (Figure \ref{fig:velocity_plots}, upper left).

This result is consistent with a picture in which warm CGM gas remains gravitationally bound across a wide range of galaxy mass
and redshift at low-$z$. \citet{Tumlinson_2011} found that \OVI\
absorbers in COS-Halos are similarly bound around $\sim L^\star$
star-forming galaxies, and \citet{Bordoloi_2014} showed the same
for cool \civ-traced gas around low-$z$ dwarf galaxies. The confinement of warm gas within the halo potential therefore
appears to be a robust feature of the low-redshift CGM across
nearly two decades in stellar mass. This picture changes
dramatically at high redshift: \citet{Bordoloi_2024} demonstrated
that even the cool CGM around high-$z$ galaxies is largely
unbound, with a significant fraction of absorbers exceeding
the escape velocity of their host halos. This redshift evolution suggests that the physical conditions regulating CGM kinematics such as halo mass, virial temperature, and feedback energy;  
conspire differently at high-$z$ to unbind gas that would
otherwise remain confined at low-$z$. The massive, high-sSFR
GOLIATH galaxies, despite their vigorous star formation and likely strong feedback, show no evidence of unbound warm gas. This suggests that their deep gravitational potential wells are sufficient to retain the \OVI-bearing material even in the presence of active outflows.

We discuss the physical origin of the multi-component kinematic
structure and the large velocity spreads observed in GOLIATH
galaxies in Section~\ref{sec:Discussion}.

\begin{figure*}
    \centering
    \includegraphics[width=1.0\linewidth]{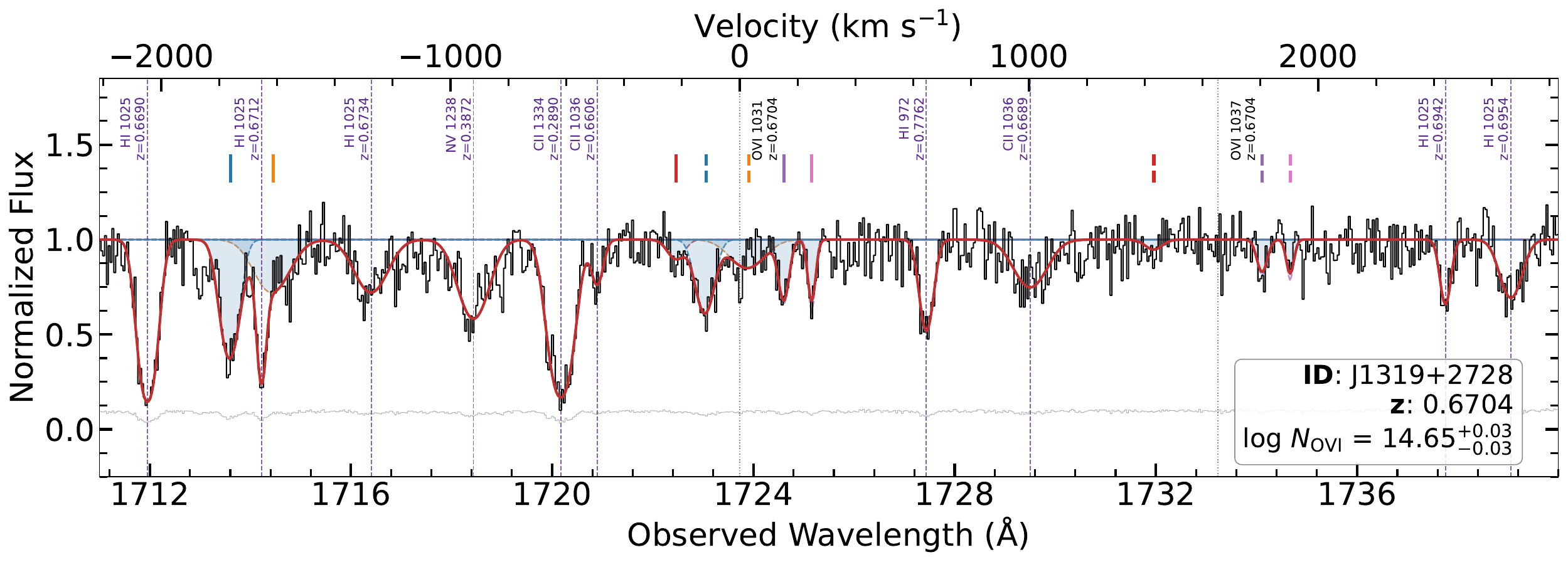}
    \caption{\textit{HST}/COS NUV spectrum of the quasar sightline J1319$+$2728 ($z = 0.6704$, $\log N_{\rm O\,VI} = 14.65^{+0.03}_{-0.03}$ integrated over $\pm 2000$ km s$^{-1}$). The velocity axis is referenced to the spectroscopic redshift of the primary target galaxy. Three distinct absorption systems are identified spanning $-2000$ to $+300$~km~s$^{-1}$, suggesting an overdense environment. The \OVI\ $\lambda 1031$ and $\lambda 1037$ components of each system are shown as solid and dashed ticks of matching color, respectively. All other ions included in the Voigt profile fit are marked with dashed lines. The reported column density corresponds to the system associated with the primary target galaxy ($-200$ to $+300$~km~s$^{-1}$). Blue shading marks the \OVI\ $\lambda 1031$ absorption regions identified in the spectrum.}
    \label{fig:CSO_spectrum}
\end{figure*}

\textit{A note on sightline J1319$+$2728.} The sightline J1319$+$2728 probe the most
massive galaxy in our sample ($\log M_\star/M_{\odot} \approx 11.9$) and its absorption profile is markedly different from the rest of the sample. As shown in Figure~\ref{fig:CSO_spectrum}, the
spectrum reveals three distinct absorption systems spanning $-2000$ to $+300$ \kms, far exceeding the $\pm1500$\kms\ search window applied to all other sightlines. SDSS imaging confirms three photometric objects
within 30\arcsec\ of the quasar sightline, suggesting an overdense environment. We associate the absorption system between -200 \kms\ and
300 \kms\ with the primary
target galaxy, which yields $\log N_{\rm O\,VI}\,[\rm cm^{-2}] =
14.01^{+0.03}_{-0.03}$; this is the value reported in
Table~\ref{tab:ovi_voigt_bms} and used in all subsequent analysis. 
The remaining two absorption systems are excluded from the kinematic statistics.

We particularly highlight J1319$+$2728 as one of the most kinematically complex \OVI\ absorption systems in any low-$z$ CGM survey. It is comparable to the landmark sightline
PG 1206$+$459 \citep{Ding_2003, Rosenwasser_2018, Tripp_2011, Churchill_1999}, which holds the record
for the strongest known CGM \OVI\ absorber (at $z<2$), with
$\log N_{\rm O\,VI}\,[\rm cm^{-2}] = 15.54 \pm 0.17$ spread over
$\gtrsim 1000$\kms, associated with a luminous
($1.3\,L^\star$) post-starburst galaxy at 68~kpc showing signatures of a recent merger. In both cases, the extraordinary kinematic complexity of absorption components spanning more than two thousand \kms\ across multiple systems; points to the role of group-scale environments and merger-driven outflows in producing the most extreme warm CGM reservoirs observed at low redshift. Two of the most kinematically rich \OVI\ systems known are both associated with massive, star-burst/post-starburst galaxies in overdense environments which suggests that large-scale structure
may be a key, and currently under-sampled, driver of \OVI\ column density at $\log M_\star/M_{\odot} \gtrsim 11$. Two of the three photometric neighbors of J1319$+$2728 lack spectroscopic redshifts; follow-up spectroscopy of this field would fully characterize the group membership and the origin of the extended absorption.

\begin{figure}[!tb]
    \centering
    \includegraphics[width=1.0\linewidth]{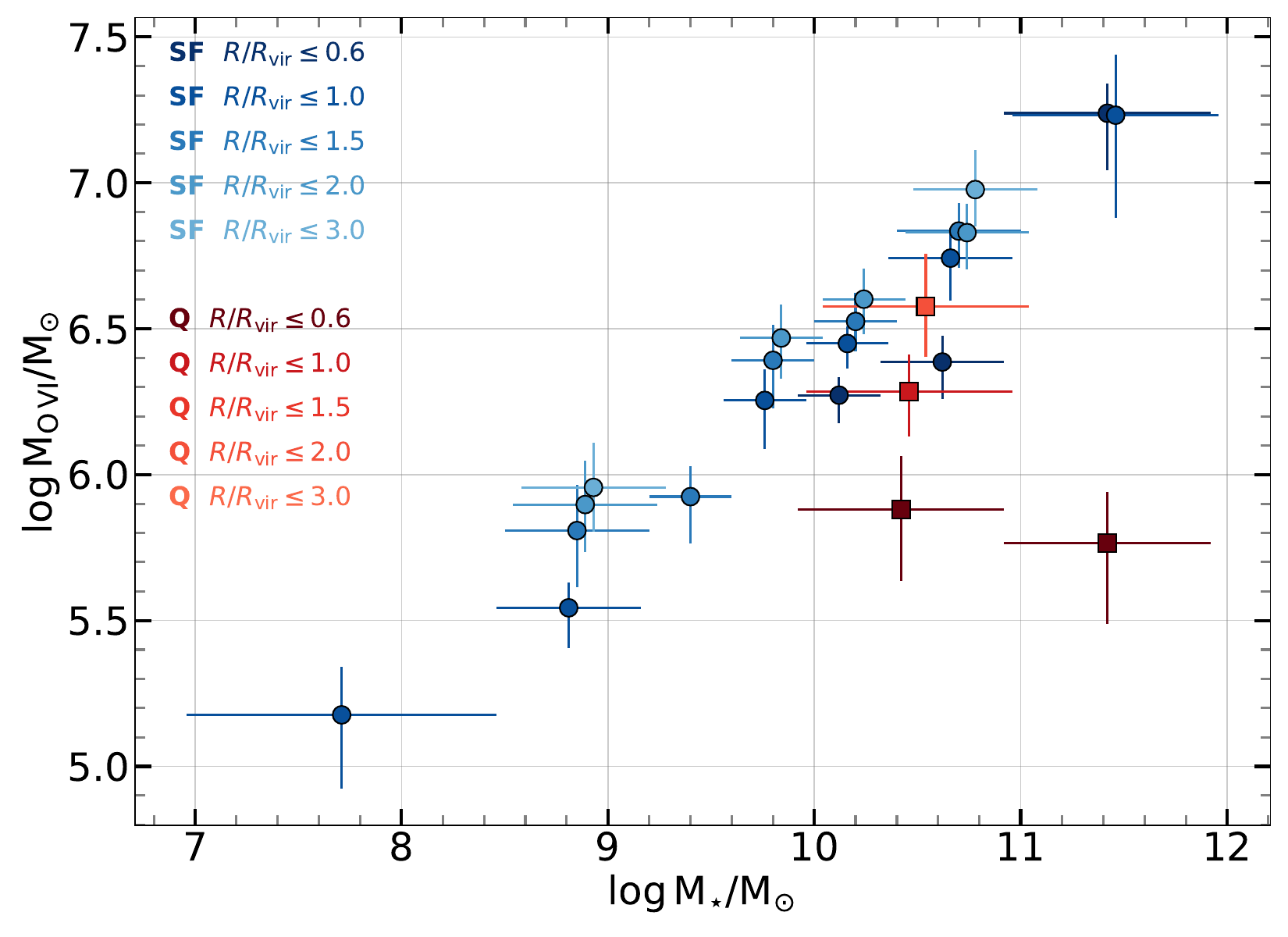}
    \caption{CGM \OVI\ mass as a function of stellar mass from the empirical annular integration method. Star-forming (blue) and quiescent (red) galaxies are shown separately; color opacity indicates the maximum virial radius integrated, from
$R/R_{\rm vir} \leq 0.6$ (darkest) to $R/R_{\rm vir} \leq 3.0$
(lightest). Points are slightly dithered in stellar mass for
clarity. At $\log M_\star/M_{\odot} \gtrsim 11$, SF galaxies show oxygen
masses $\sim 1.5$~dex higher than quiescent galaxies of the
same stellar mass, contrary to the expectation from virial
thermometer models that predict
suppressed \OVI\ at this mass range. A threshold of
$\log N_{\rm O\,VI, Threshold}\,[\mathrm{cm}^{-2}] = 14.0$ is used for the covering fraction calculation.}
    \label{fig:O_mass}
\end{figure}

\begin{figure}[!tb]
    \centering
    \includegraphics[width=1.0\linewidth]{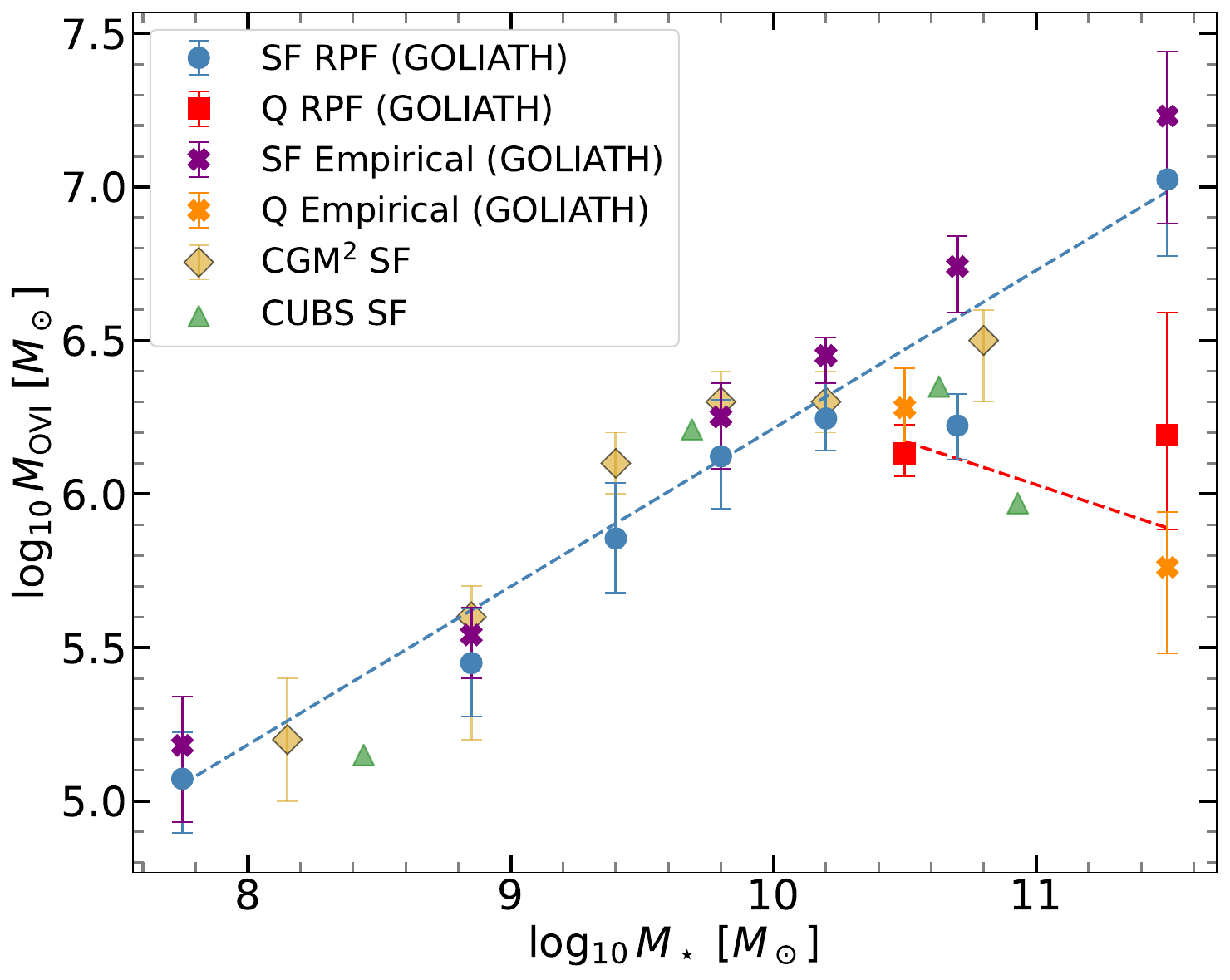}
    \caption{CGM \OVI\ mass as a function of stellar mass, estimated from the radial profile fitting (RPF; circles and squares) and empirical annular integration (stars) methods. Star-forming (SF) and quiescent (Q) galaxies are shown in cool and warm colors, respectively. Archival comparison samples include CUBS \citep{Qu_2024}, and CGM$^2$ \citep{Tchernyshyov_2022}. Dashed lines show power law fits to the SF (purple) and quiescent (orange) empirical estimates. Both methods are in good agreement across all mass bins. At $\log M_\star/M_{\odot} \gtrsim 11$, the SF and quiescent oxygen mass estimates diverge, with SF galaxies continuing to rise while quiescent galaxies turn over, reflecting the sharp decline in \OVI\ detections at high mass in passive systems.}
    \label{fig:rpf_o_mass}
\end{figure}

% Potential change: Find table of O ionization fraction as change in temperature and stellar mass

\subsection{Total \OVI\ Mass in the CGM}\label{sec:oxygen_mass}

To quantify the total \OVI\ content of galaxies' CGM across the SF--quiescent divide, we estimate the total CGM \OVI\ mass as a function of stellar mass with two methods. 
Both methods share the same base definition,
\begin{equation}
    M_{\rm O\,VI}(R) = \pi R^2 m_{\rm O}\langle N_{\rm O\,VI}\rangle(R)
    \label{eq:ovi_mass_eq}
\end{equation}

where $\langle N_{\rm O\,VI} \rangle (R)$ is the mean column density
at radius $R$ and $m_{\rm O}$ is the oxygen atom mass. 

It is possible to convert this mass to the total warm-oxygen by dividing equation \ref{eq:ovi_mass_eq} by an ionization correction factor $f_{\rm O\,VI}$. 

%We assume collisional ionization equilibrium with , maximum ionization fraction for \OVI\ \citep{Gnat_Sternberg_2007}, making all mass estimates conservative lower limits.

\subsubsection{Empirical Method}
Following \citet{Tumlinson_2011} and \citet{Bordoloi_2014}, we
divide each population into stellar mass bins and compute the \OVI\ mass by integrating in physical impact parameter.
Unlike previous work, which bins sightlines directly in physical
separation $R$, we first select and bin sightlines in
$R/R_{\rm vir}$ to ensure equivalent halo-scaled radial coverage
across the stellar mass range. The mean virial radius of each
bin is then used to convert the annular boundaries back to
physical kpc for the mass integration. Within each annulus the
column density is weighted by the covering fraction ($f_c$),
\begin{equation}
    \langle N_{\rm O\,VI}\rangle_i = f_c(r)
    \langle N_{\rm O\,VI}\rangle_{\rm annulus}.
\end{equation}
Each annulus is grown in steps of $0.1\,R/R_{\rm vir}$ until
it contains at least eight sightlines or reaches the maximum
integration radius, minimizing uncertainty in the covering
fraction. This method is grounded directly in the observed
\OVI\ without interpolation, but is sensitive to sparse
sightline coverage at small $R/R_{\rm vir}$. We label any \OVI\ mass estimated using this method as the empirical mass.

\subsubsection{Radial Profile Fitting Method}
Following \citet{Tchernyshyov_2022}, we
model the \OVI\ column density radial profile with a beta
function,
\begin{equation}
    f(x\mid\theta) = \log_{10} \left[ N_0 \left(1 +
    \left(\frac{x}{r_c}\right)^{\!2}\right)^{-\beta}
    + N_b\right],
\end{equation}
where $x = R/R_{\rm vir}$, $N_0$ is the central column density,
$r_c$ is the core radius, $\beta$ sets the slope, and $N_b$
is a fixed IGM background term ($\log N_b\,[\mathrm{cm}^{-2}]\sim 12.5$; \citealt{Nelson18})
The background term is included in the fit but excluded from
the mass integration. The oxygen mass is then computed from
the area-weighted mean column density within the virial radius,
\begin{equation}
    \langle N_{\rm O\,VI}\rangle_{r_{\rm vir}} =
    \frac{2}{R^2_{\rm vir}} \int_0^{R_{\rm vir}}
    N_{\rm O\,VI}(R')\,R'\,dR'.
\end{equation}
We fit the model using an MCMC sampler \citep{karamanis2021zeus}
with the following negative log-likelihood, treating
non-detections as censored upper limits via a complementary
error function,
\begin{multline}
    -\ln\mathcal{L}_{\rm det} = \frac{1}{2}\sum_i
    \left[\frac{\Delta_i^2}{\sigma_i^2} +
    \ln\sigma_i^2\right], \\
    \ln\mathcal{L}_{\rm lim} = \sum_j \ln\left[
    \frac{1}{2}\left(1 + {\rm erf}\!\left(
    \frac{\Delta_{{\rm lim},j}}{\sqrt{2}\,
    \sigma_{{\rm lim},j}}\right)\right)\right].
\end{multline}

We label the mass estimates from this method as \OVI\ mass from the radial profile fitting (RPF) method.

\subsubsection{\OVI\ Mass Estimates}\label{sec:o_mass_results}

Figures~\ref{fig:O_mass} and \ref{fig:rpf_o_mass} show the CGM
\OVI\ mass estimates from the empirical and RPF methods,
respectively. Both methods are in good agreement across all
stellar mass bins, differing by at most $\sim$0.5~dex,
confirming that two independent approaches yield a consistent
picture of the \OVI\ budget.

From the empirical method (Figure~\ref{fig:O_mass}), the SF CGM
CGM \OVI\ mass rises steadily with stellar mass across
$\log M_\star/M_{\odot}\sim 7.5$--$11.5$, with deeper integration radii contributing additional mass as expected. 
The quiescent population is systematically offset to lower \OVI\ masses at all stellar masses. The contrast is sharpest at
$\log M_\star /M_{\odot} \gtrsim 11$, where SF galaxies reach
$\log M_{\rm O\,VI}/M_\odot \sim 7.25\pm0.25$ while quiescent galaxies of
the same stellar mass show $\log M_O/M_\odot \sim 5.75\pm0.25$,
a difference of $\sim 1.5$~dex.

Figure~\ref{fig:rpf_o_mass} shows the RPF \OVI\ mass
estimates alongside literature measurements from CUBS
\citep{Qu_2024} and CGM$^2$
\citep{Tchernyshyov_2022}. Within the overlapping mass
range ($\log M_\star / M_{\odot} \sim 8$--$11$), our SF and quiescent
estimates are in good agreement with both surveys, validating
the RPF method. The GOLIATH data points extend these relations
into the $\log M_\star/ M_{\odot} \gtrsim 11$ regime not previously
sampled by either survey. At $\log M_\star/M_{\odot} \sim 11.5$, GOLIATH provides the first estimates of the \OVI\ mass for both massive SF galaxies and quiescent galaxies: the SF RPF point
reaches $\log M_{\rm O\,VI}/M_\odot \sim7\pm0.25$, while the quiescent
point sits at $\log M_{\rm O\,VI}/M_\odot \sim6.2\pm0.4$. Both populations
are well described by power laws of the form
\begin{equation}
    \log M_{\rm O\,VI} = \left(\alpha\,\log M_\star +
    \gamma\right)
\end{equation}
where for SF galaxies $\alpha = 0.52\pm$
0.04 and $\gamma = 1.06\pm$
0.40, and for quiescent galaxies
$\alpha = -0.28$ and
$\gamma = 9.13$. The opposing
slopes with SF rising, quiescent declining and the
$\sim 1.1$~dex divergence at $\log M_\star/M_{\odot} \sim 11.5$
demonstrate that star formation activity, not halo mass,
is the dominant regulator of the warm oxygen reservoir
at this mass scale, inconsistent with the virial
thermometer model \citep{Oppenheimer_2016}.

The excess \OVI\ mass in massive SF galaxies is
consistent with feedback-driven radiative shocks actively
maintaining the warm CGM \citep{Heckman_2002, Bordoloi_2017}. In this
scenario, outflows shock-heats the surrounding gas, which subsequently cools radiatively through the \OVI\ temperature window at $T \sim 10^{5.5}$~K, sustaining the elevated column densities and oxygen masses that are observed. As these galaxies quench and feedback subsides, this replenishment mechanism shuts off, the warm gas is heated to higher ionization states or consumed, and the oxygen mass drops toward the quiescent locus. The multi-component kinematic
structure of the GOLIATH absorbers
(Section~\ref{sec:kinematics}) is consistent with such an
active feedback environment.

\section{Discussion}\label{sec:Discussion}

\begin{figure}[!tb]
    \centering
    \includegraphics[width=1.0\linewidth]{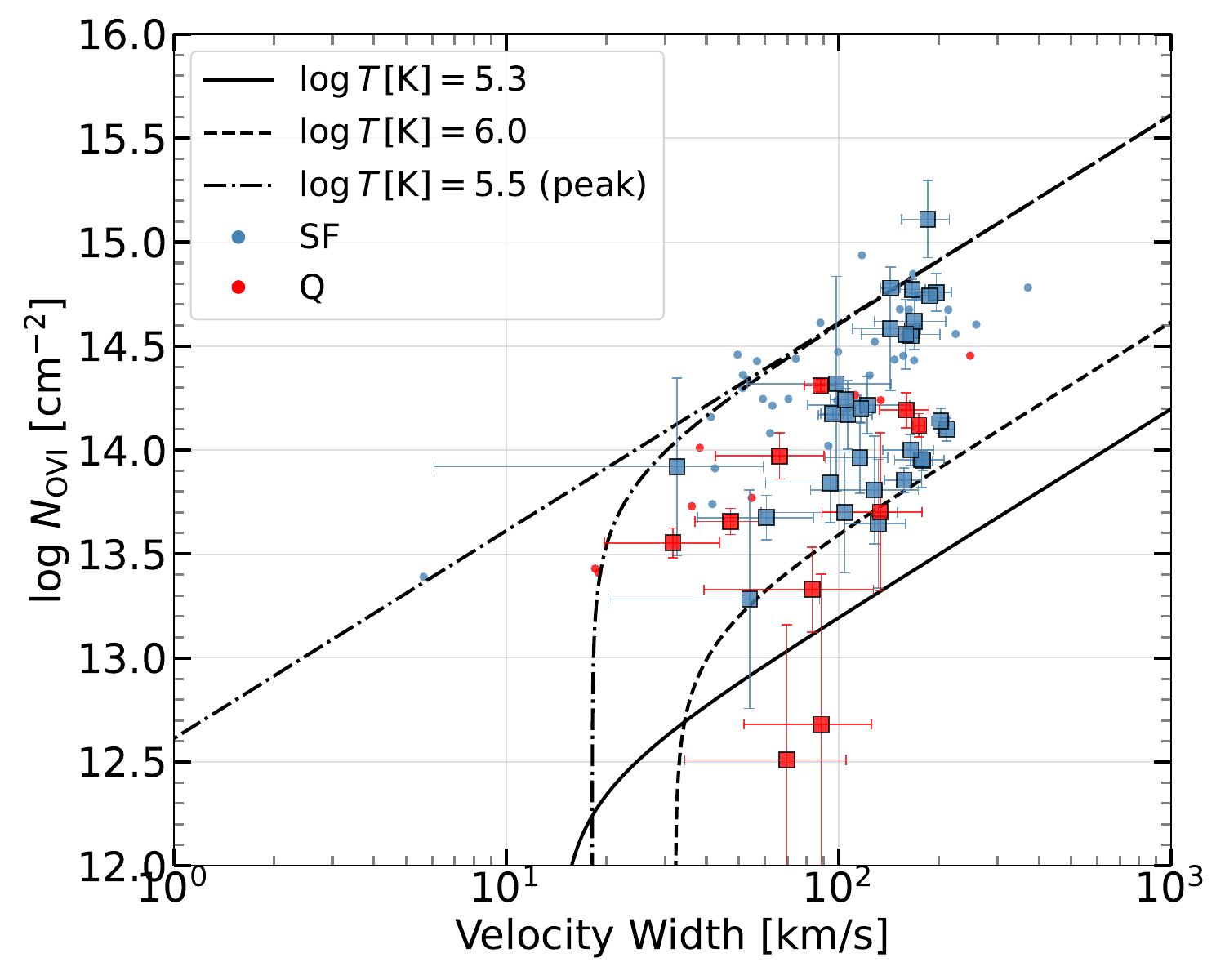}
    \caption{\OVI\ column density of individual absorption components
    as a function of absorption line width for GOLIATH (squares) and COS-Halos (circles) components, separated into star-forming (blue) and quiescent (red) populations. Curves show the predicted \OVI\ column density in radiatively cooling post-shock gas model  \citep{Bordoloi_2017} assuming
    collisional ionization equilibrium (CIE),
    at three characteristic temperatures: $\log T = 5.3$ K (solid), $\log T \approx 5.5$ K corresponding to the peak \OVI\ ionization fraction (dash-dot), and $\log T = 6.0$ K (dashed). Star-forming components are broadly consistent with gas cooling radiatively
    through the \OVI\ peak ionization temperature, while quiescent components scatter toward the hotter $\log T = 6.0$ K curve, consistent with gas being heated above the \OVI\ window by the higher virial temperatures of their host halos.}
    \label{fig:cooling_flows}
\end{figure}

The GOLIATH survey targets the stellar mass regime ($\log M_\star/M_{\odot} \gtrsim 11$) where ``virial thermometer" models predict a sharp decline in \OVI\ absorption as the halo virial temperature rises above the \OVI\ peak
ionization temperature \citep{Gnat_Sternberg_2007, Oppenheimer_2016}. Our observations reveal that massive SF galaxies maintain a high \OVI\ covering fraction of
$\approx$62.5\%, a CGM \OVI\ mass $\sim 1.1$~dex above their quiescent counterparts, and complex multi-component kinematic structure. The data are inconsistent with virial thermalization alone as the dominant regulator of the warm CGM in actively star-forming massive halos.

\subsection{CGM Cooling}
At $\log M_\star /M_{\odot} \gtrsim 11$, the halo virial temperature exceeds $T_{\rm vir} \sim 10^{6}$~K, well above the \OVI\ peak ionization temperature of $T \sim 10^{5.5}$~K \citep{Gnat_Sternberg_2007}. The virial thermometer model
\citep{Oppenheimer_2016} predicts that \OVI\ should be
collisionally ionized away at this mass scale. This prediction is consistent with the quiescent population in our sample: the covering fraction drops to $\approx$ 24.0\% and the \OVI\ declines at $\log M_\star/M_{\odot} \gtrsim 11$, with the few \OVI\ detections in quiescent systems scattering toward the hotter $\log T = 6.0 $K cooling curve in Figure~\ref{fig:cooling_flows}, consistent with gas residing at the high-temperature end of the \OVI\ ionization window. The virial thermometer model therefore remains a viable explanation for the residual \OVI\ seen in passive galaxies at this mass scale.

For the massive SF population, however, the data are inconsistent with this picture. The GOLIATH SF components cluster along the $\log T \approx 5.5$ peak ionization curve in Figure~\ref{fig:cooling_flows}, consistent with a feedback-driven shock model in which outflows shock-heat the surrounding CGM and the gas subsequently cools radiatively through the \OVI\ temperature window \citep{Heckman_2002, Bordoloi_2017}. In this scenario, feedback-driven outflows, continuously replenish the warm gas reservoir as gas cools behind a shock front radiatively. They form fresh \OVI\ gas in the halo sustaining the high covering fractions
and oxygen masses we observe despite the elevated virial temperature of these halos. The radiative cooling time of gas at $T \sim 10^{5.5}$~K and CGM densities ($n_H \sim 10^{-4}$~cm$^{-3}$) is short, $t_{\rm cool} \sim 10$--$100$~Myr \citep{Gnat_Sternberg_2007}, meaning that without active replenishment the \OVI\ phase would rapidly cool out of the observable temperature window. The persistence of high \OVI\ column densities in the GOLIATH SF sample therefore requires a sustained source of warm gas, consistent with ongoing feedback rather than a leftover relic population.

An alternative explanation is that the observed \OVI\ is part of a relic population which is actively recombining to lower ionization states. This is disfavored on two grounds. First, the recombination
timescale at CGM densities ($t_{\rm rec} \sim R_{\rm vir} / \sigma_v \sim 1$~Gyr) is an order of magnitude longer than $t_{\rm cool}$, meaning any relic gas would cool out of the \OVI\ window long before it could recombine
\citep{Oppenheimer_2016, Zahedy_2019_COS_LRG}. Second, relic gas cooling from above the \OVI\ window would populate the $\log T \,\rm[K]= 6.0$ curve in Figure~\ref{fig:cooling_flows} rather than the observed peak ionization region at $\log T \,[\rm K] \approx 5.5$.

As galaxies quench and feedback subsides, the radiative cooling model predicts that the \OVI\ phase should disappear on the cooling timescale of
$t_{\rm cool} \sim 10$--$100$~Myr; far shorter than both the recombination timescale and the $\sim 1$~Gyr SF-to-quiescent transition \citep{Schawinski_2014}. As gas cools below $T \sim 10^{5.5}$~K into lower ionization states, the \OVI\ reservoir is no longer replenished by feedback; thus, the covering fraction and warm oxygen mass drop rapidly toward the quiescent locus. The absence of GOLIATH galaxies in the green valley
($-12 \lesssim \log\,{\rm sSFR}\,[\rm yr^{-1}] \lesssim -11$) is consistent with this
rapid transition: the \OVI\ depletion timescale is short
enough that galaxies passing through the green valley
are unlikely to be caught with a full warm gas reservoir,
and we observe these systems either before or after
the depletion event, not during it.

\subsection{\OVI\ Gas Kinematics}
The GOLIATH absorbers show a mean of 3 kinematic components per sightline at $\log M_\star/M_{\odot} \gtrsim 11$, higher than the 2 components seen at lower masses in COS-Halos and MUSEQuBES (Section~\ref{sec:kinematics}). Because each sightline probes only a single pencil beam through the CGM, the multiple components likely trace gas at different physical locations and origins within the halo and infalling
streams, outflowing shells, and stripped group gas may
all contribute \citep{Ho_2026}. While turbulent motions could contribute to
the kinematic broadening \citep{Qu_2026}, the S/N of
our spectra is insufficient to fit a turbulence broadening
parameter independently. The extreme case of J1319$+$2728 with absorption spanning $-2000$ to $+300$~km~s$^{-1}$
across three distinct systems illustrates how group
environments can amplify kinematic complexity along a
single sightline (Section~\ref{sec:kinematics}). Despite
this complexity, all detected \OVI\ components remain
gravitationally bound ($|v/v_{\rm esc}| < 1$), consistent
with COS-Halos \citep{Tumlinson_2011} and COS-Dwarfs
\citep{Bordoloi_2014} at low redshift. This contrasts with high-$z$ galaxies where even cool CGM gas is largely unbound \citep{Rudie_2019, Bordoloi_2024}, suggesting that the deeper gravitational potential wells of low-$z$ massive halos retain the warm gas even in the presence of vigorous feedback.

\subsection{Implications for Theory }
While simulations have not directly explored the CGMs of high-mass star-forming galaxies, we can infer some results in regard for GOLIATH systems.  \citet{Oppenheimer_2016} shows that the \OVI\ mass inside 150 kpc peaks at $\log M_{\rm halo}/M_\odot\approx 12$ in EAGLE-CGM zoom simulations, but steadily declines at higher halo mass, even for moderately star-forming massive galaxies, in contrast to GOLIATH systems. In the IllustrisTNG100 simulation, \citet{Nelson18} finds strong \OVI\ associated with star-forming galaxies and weak \OVI\ at higher stellar and halo mass when AGN feedback quenches the galaxy and removes material from the CGM. They do demonstrate that more passive galaxies at fixed stellar mass have less \OVI, but GOLIATH systems are not directly modeled.  Similarly, \citet{Appleby2023} also finds this trend in SIMBA simulations, reproducing the observed dichotomy of \OVI\ strength for star-forming and passive galaxies at fixed stellar mass, but they also do not model high-mass star-forming  GOLIATH-like systems.  

In observations, while estimating the minimum total Oxygen mass in the CGM an \OVI\ ionization fraction of $f_{\rm O\,VI}=0.2$ is often used \citep{Tumlinson_2011}. However, simulations predict a much lower global $f_{\rm O\,VI}$. Both EAGLE-CGM zooms and IllustrisTNG100 volume have $f_{\rm O\,VI}\approx 0.03$ at $\log M_{\rm halo}/M_\odot \approx 12$, and decline below $0.01$ by $\log M_{\rm halo}/M_\odot \approx13$ in the extended CGM counting bound gas beyond the virial radius.  \citet{wijers2020} finds $f_{\rm O\,VI}\approx0.01$ at $\log M_{\rm halo}/M_\odot \approx 12$ in the main EAGLE (100 Mpc)$^3$ volume, but their predicted column densities are too low compared to observations.  We therefore restrict our analysis to report only total \OVI\ mass of the CGM which will allow future theory work to directly compare with these measurements. 

Another aspect is the calculation of the total oxygen budget nucleosynthesized and released by stars.  This has been explored in \citet{Peeples_2014} and \citet{Oppenheimer_2016}, with the total oxygen yield being 2.5-3\% of the stellar mass and a majority of this oxygen being released from the galaxy into the CGM and beyond.  Other simulations, including FIRE have lower oxygen yields \citep{Muratov_2017}, as supernova yields inputted into these simulations have significant uncertainty.  A simple calculation of the $\log M_{\rm O\,VI}/M_\odot=7.3$ calculated for GOLIATH systems with an average $M_{\star}=10^{11.4}\ M_{\odot}$ from Figure 9 can be achieved with $f_{\rm O\,VI}\approx 0.01$ and an oxygen yield of $\approx$1\%, which are both in the realm of simulation results.  The difference for GOLIATH systems is that $f_{\rm O\,VI}$ is significantly higher than passive galaxies, unless passive gaseous halos are significantly more  evacuated (i.e. on the order of 1.1 dex more) than GOLIATH systems.

Another possibility is the need for higher oxygen yields and/or more efficient metal transport by feedback.  \citet{piacitelli2026sos} demonstrate that dwarf galaxy simulations (FIRE-2 and M+M) systematically under-predict CGM oxygen content.  They find higher $f_{\rm O\,VI}\approx0.1$ in dwarf galaxy ($\log M_\star/M_\odot= 7.7-8.7$) CGMs, which is consistent with the trend of the above simulations that achieve higher $f_{\rm O\,VI}$ at lower mass where photo-ionized \OVI\ can reach $f_{\rm O\,VI}>0.2$.  Nevertheless, they argue that a \OVI\ shortfall still exists, but it is not clear if this is a problem that extends to higher mass GOLIATH-like systems.

\subsection{Future Work}
Observations of higher ionization tracers including \ion{Ne}{8},
\ion{C}{4}, and \ion{N}{5} would distinguish between collisional and photoionization channels and constrain the temperature structure of the multiphase CGM at this mass scale
\citep{Garza_2025}.
Further characterization of the cool and hot CGM of these systems would allow for metallicity and total baryon budget calculation -- filling a major gap in the literature \citep{Tumlinson_2017}.

\section{Conclusions and Summary}
We present the first results from the GOLIATH survey, an HST/COS survey of the multiphase CGM of massive ($\log M_\star/M_{\odot} \gtrsim 11$), blue, starburst and post-starburst galaxies. Our results demonstrate that the inner CGM of massive, actively star-forming galaxies harbors a rich reservoir of warm \OVI\ traced gas that persists despite these massive halos existing in virial temperatures well above the \OVI\ peak ionization temperature. This result is inconsistent with the virial thermometer as a complete description of warm gas regulation at the high-mass end of the star-forming sequence and points instead to feedback-driven radiative cooling as the dominant mechanism sustaining the warm CGM in these systems. Our principal conclusions are as follows.

\begin{enumerate}

\item \textit{The warm CGM of massive SF galaxies is not suppressed by virial heating.}
    At $\log M_\star/M_{\odot} \gtrsim 11$, where virial thermalization models predict a sharp decline in \OVI\ abundance \citep{Oppenheimer_2016}, massive SF galaxies maintain an \OVI\ covering fraction of 62.5$\pm$14\% at $\log N_{\rm O\,VI} \geq 14.0$ and $R/R_{\rm vir} \leq 0.6$, compared to 24$\pm$10\% for quiescent galaxies of the same stellar mass. The full GOLIATH sample has a covering fraction of 64.7$\pm$11\%. The \OVI\ column density scatter of $\sigma= 0.481$~dex about the SF power law is 58\% larger than the $\sigma = 0.304$~dex scatter of lower-mass SF galaxies. This  indicates that the warm CGM spans a wide range of states at this mass scale, from fully replenished to partially depleted. Halo mass alone cannot account for the observed SF--quiescent dichotomy. A Fisher's exact test confirms that star-forming galaxies show detection 4.5 times more than quiescent galaxies for a threshold $\log N_{Th}=14.0$ (p $<$ 0.05).
    
    \item \textit{\OVI\ column density increases with both stellar mass and sSFR.} In star-forming galaxies, \OVI\ column density rises by nearly 1~dex from $\log M_\star/M_{\odot} \sim 8$ to $\sim 11.5$, with quiescent galaxies offset $\sim 0.5$~dex below SF galaxies at fixed stellar mass in the $\log M_\star/M_{\odot} = [11,12]$ inner-CGM ($R/R_{\rm vir}<0.6$) bin. \OVI\ column density similarly increases with sSFR, with GOLIATH galaxies occupying the upper locus of the SF branch at $\log\,{\rm sSFR} [\rm yr^{-1}] \gtrsim -10$, including several sightlines among the highest \OVI\ column densities in any low-$z$ CGM survey at $\log N=$15.3 and 15.2 for J0745+1919 and PG1206\text{+}459, respectively. We confirm a monotonic increase between the column density and stellar mass (Generalized Kendall's Tau test $\tau=0.193,\,p<0.0001$). We are also able to confirm that inner-CGM ($R/R_{\rm vir}<0.6$) O VI column density is monotonically correlated to the sSFR of galaxies ($\tau=0.393,~p<0.0001$).

    \item \textit{The warm gas is consistent with feedback-driven radiative cooling.} The total \OVI\ absorption line widths and column densities of individual GOLIATH SF components are consistent with the scenario of \OVI\ being created by gas cooling radiatively behind feedback driven shocks \citep{Bordoloi_2017}, clustering along the $\log T~[K] \approx 5.5$ peak \OVI\ ionization curve (Figure~\ref{fig:cooling_flows}). The radiative cooling time at $T \sim 10^{5.5}$~K and CGM densities ($n_H \sim 10^{-4}$~cm$^{-3}$) is $t_{\rm cool} \sim 10$--$100$~Myr \citep{Gnat_Sternberg_2007}, meaning the \OVI\ phase must be continuously replenished by feedback-driven outflows to persist. The quiescent components scatter toward the hotter $\log T \,[\rm K]= 6.0$ curve, consistent with gas that has been heated above the \OVI\ window as feedback subsides, which the virial thermometer model can accommodate for passive systems \citep{Oppenheimer_2016}.

    \item \textit{\OVI\ is a sensitive tracer of the SF-quiescent transition.} The short cooling time $t_{\rm cool} \sim 10$--$100$~Myr makes \OVI\ a sensitive indicator of quenching: once feedback ceases, the warm gas reservoir depletes on timescales an order of magnitude shorter than the $\sim 1$~Gyr SF-to-quiescent transition \citep{Schawinski_2014} and far shorter than the recombination timescale of $t_{\rm rec} \sim R_{\rm vir}/\sigma_v \sim 1$~Gyr at CGM densities. The \OVI\ phase therefore acts as a close tracer of feedback activity: it is present when feedback is active, and absent once it stops, well before the galaxy itself crosses the SF--quiescent divide. The $\sim 1.1$~dex SF--quiescent offset in CGM \OVI\ mass at $\log M_\star/M_{\odot} \sim 11.5$ directly reflects this rapid depletion, making \OVI\ a uniquely sensitive probe of the quenching process at the high-mass end.

    \item \textit{The SF--quiescent \OVI\ mass offset grows with stellar mass.}
    Both the empirical annular integration and radial profile fitting methods yield consistent CGM \OVI\ estimates, extending the census for star-forming galaxies to $\log M_\star/M_{\odot} \sim 11.5$. This reaches the highest seen \OVI\ mass at $\log M_{\text O\,VI} = 7.25\pm 0.25$ . The SF warm-oxygen mass follows a rising power law ($\alpha = 0.51\pm0.04$ , $\gamma = 1.06\pm0.40$ ) while the quiescent population follows a declining relation ($\alpha
    = -0.28$, $\gamma =
    9.13$), producing a $\sim 1.1$~dex divergence at $\log M_\star/M_{\odot} \sim 11.5$. These are the first \OVI\ mass estimates at this stellar mass for both populations. 
    \item \textit{The warm gas is kinematically complex
    but gravitationally bound.}
    The gas kinematics reveal that massive galaxies have, on average, three velocity components.
    The velocity centroid distribution across the full
    sample is well described by a Gaussian centered at
    $0.01\,v_{\rm esc}$ with $\sigma = 0.08\,v_{\rm
    esc}$, confirming that all detected components
    remain gravitationally bound ($|v/v_{\rm esc}| \leq  
    1$). This is consistent with the low-$z$ CGM across
    two decades in stellar mass \citep{Tumlinson_2011,
    Bordoloi_2014} and contrasts with the largely
    unbound cool CGM at high redshift
    \citep{Bordoloi_2024}.
    \item \textit{GOLIATH systems are not represented in current CGM simulations.}
    We can infer some results in GOLIATH systems from \citet{Oppenheimer_2016, Nelson18}; and \citet{Appleby2023} simulations which predict that the \OVI\ ionization fraction should be much lower ($f_{\rm O\,VI}\approx0.01-0.03$) than the peak ionization fraction at $\log \,T\,[\rm K]\approx5.5$ ($f_{\rm O\,VI}=0.2$).
    The GOLIATH sample's \OVI\ mass ($\log M_{\rm O\,VI}/M_\odot=7.3$) can be achieved with a $f_{\rm O\,VI}\approx0.01$ and an oxygen yield of $\approx1\%$ which are within the realm of possibilities for simulations. This required that simulations are able to replicate the significantly higher $f_{\rm O\,VI}$ in star-forming galaxies compared to passive gaseous halos. 
\end{enumerate}
The GOLIATH survey establishes that star formation
activity sustains a warm oxygen reservoir in massive
galaxy halos that would otherwise be thermally suppressed,
and that this reservoir depletes rapidly once feedback
ceases, making \OVI\ a uniquely powerful tracer of the
quenching process. Current cosmological simulations
systematically under-predict this oxygen budget
\citep{piacitelli2026sos},
pointing to incomplete feedback prescriptions at the
high-mass end. Future papers from this survey will
characterize the full multiphase CGM structure around
these galaxies through observations of cooler tracers
(\ion{H}{1}, \ion{C}{2}, \ion{O}{1}, \ion{Si}{3}) and other warm-
hot tracers (\ion{C}{4}, \ion{N}{5}, \ion{Ne}{8}),
building a complete census of the baryonic content of
massive galaxy halos across the SF--quiescent transition
\citep{Garza_2025}.

\begin{acknowledgments}

This work is based on observations with the NASA/ESA
Hubble Space Telescope through program number HST-GO17211, obtained at the Space Telescope Science Institute, which is operated by the Association of Universities for Research in Astronomy, Incorporated, under NASA contract NAS5-26555.

Some of the data presented herein were obtained at Keck Observatory, which is a private 501(c)3 non-profit organization operated as a scientific partnership among the California Institute of Technology, the University of California, and the National Aeronautics and Space Administration. The Observatory was made possible by the generous financial support of the W. M. Keck Foundation. The authors wish to recognize and acknowledge the very significant cultural role and reverence that the summit of Mauna Kea has always had within the Native Hawaiian community. We are most fortunate to have the opportunity to conduct observations from this mountain.

%LBT acknowledgement
The LBT is an international collaboration among institutions in the United States and Europe. At the time data were acquired for this research, LBT Corporation Members were the University of Arizona on behalf of the Arizona Board of Regents; Istituto Nazionale di Astrofisica, Italy; LBT Beteiligungsgesellschaft, Germany, representing the Max-Planck Society, the Leibniz Institute for Astrophysics Potsdam, and Heidelberg University; and The Ohio State University, representing The Ohio State University, University of Notre Dame, University of Minnesota, and University of Virginia.  This research used the facilities of the Italian Center for Astronomical Archives (IA2) operated by INAF at the Astronomical Observatory of Trieste. Observations have benefited from the use of ALTA Center (alta.arcetri.inaf.it) forecasts performed with the Astro-Meso-Nh model. Initialization data of the ALTA automatic forecast system come from the General Circulation Model (HRES) of the European Centre for Medium Range Weather Forecasts.

We thank T. Treu, W. Sheu, H. Paugnat, S. Knabel, D. Williams, and P. Mozumdar for providing one of the Keck/KCWI observations used to measure a GOLIATH galaxy redshift.

\end{acknowledgments}

\facilities{HST(COS), KECK(KCWI/KCRM)}

\software{rbcodes \citep{bordoloi_2025_15723701}, 
          rbvfit \citep{bordoloi_2025_rbvfit}, 
          Prospector \citep{Prospector_2021},
          Astropy \citep{astropy:2013, astropy:2018, astropy:2022}}

\clearpage
\bibliography{GOLIATH}{}
\bibliographystyle{aasjournalv7}

\clearpage
\appendix
\vspace*{-30pt}
% --- High resolution figure (3 pages) ---
\begin{figure}[tbhp]
    \centering
    \includegraphics[page=1, width=0.9\textwidth]{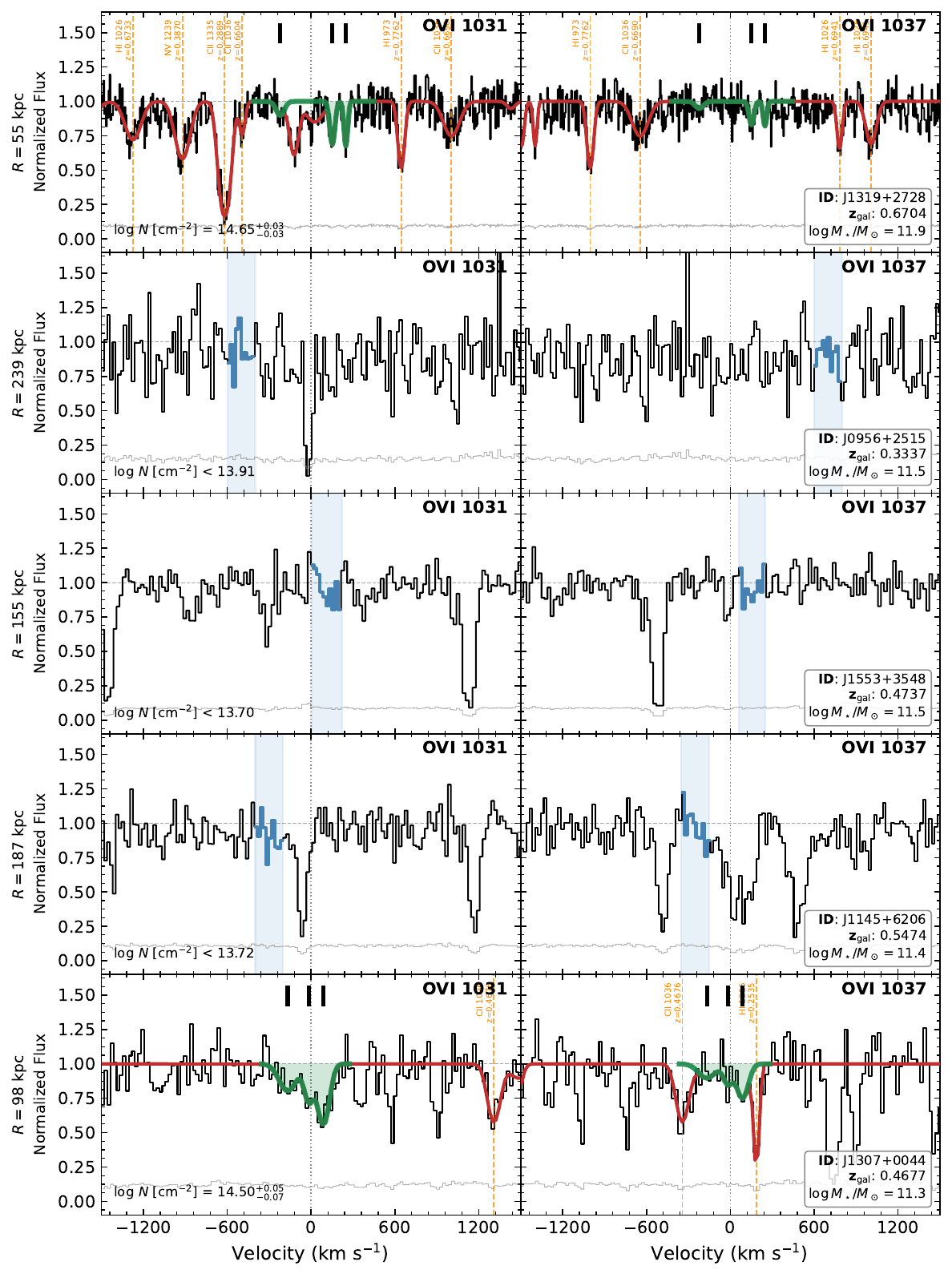}
\end{figure}
\clearpage
\begin{figure}[p]
    \ContinuedFloat
    \centering
    \includegraphics[page=2, width=\textwidth]{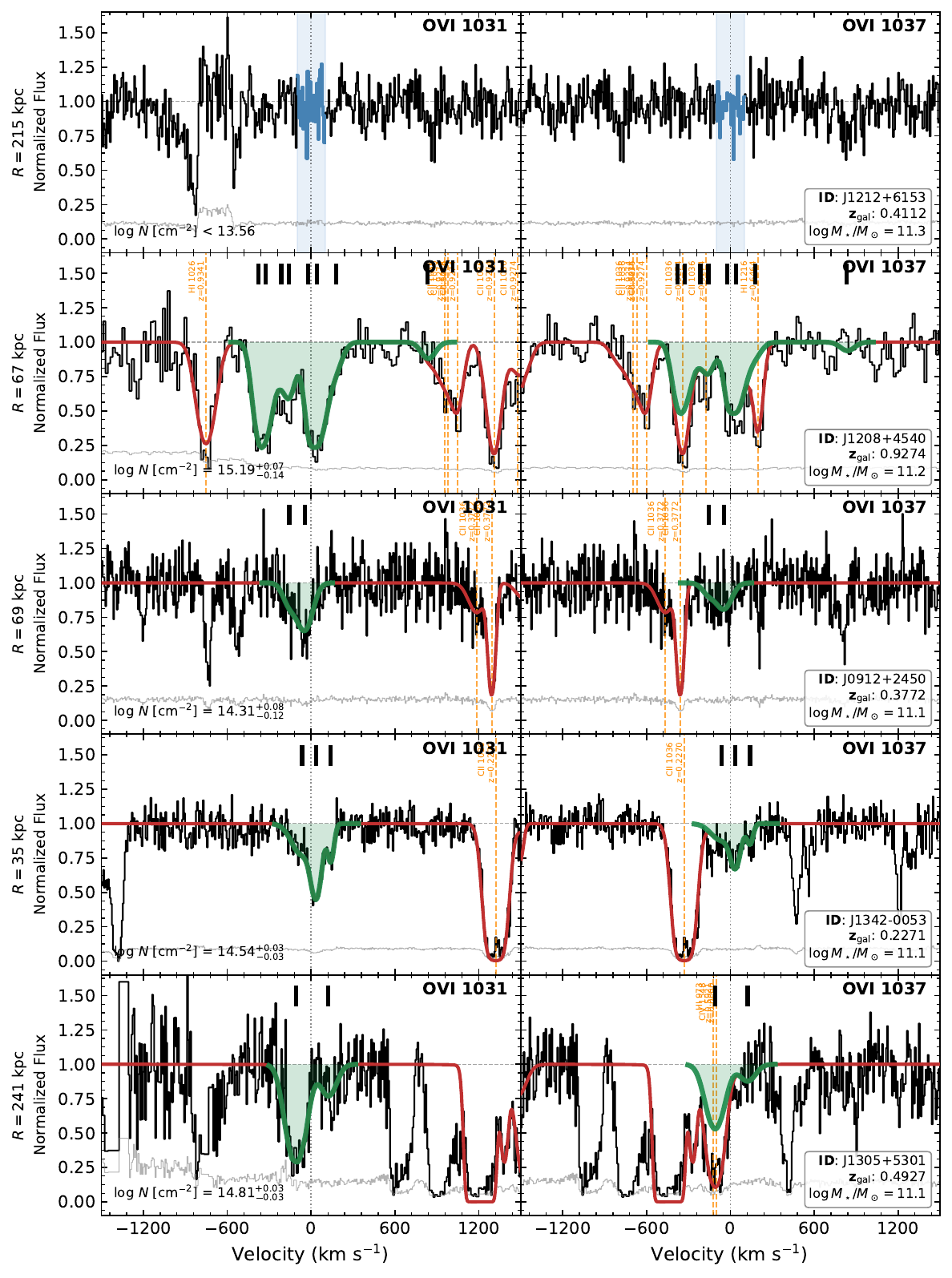}
\end{figure}
\clearpage
\begin{figure}[p]
    \ContinuedFloat
    \centering
    \includegraphics[page=3, width=\textwidth]{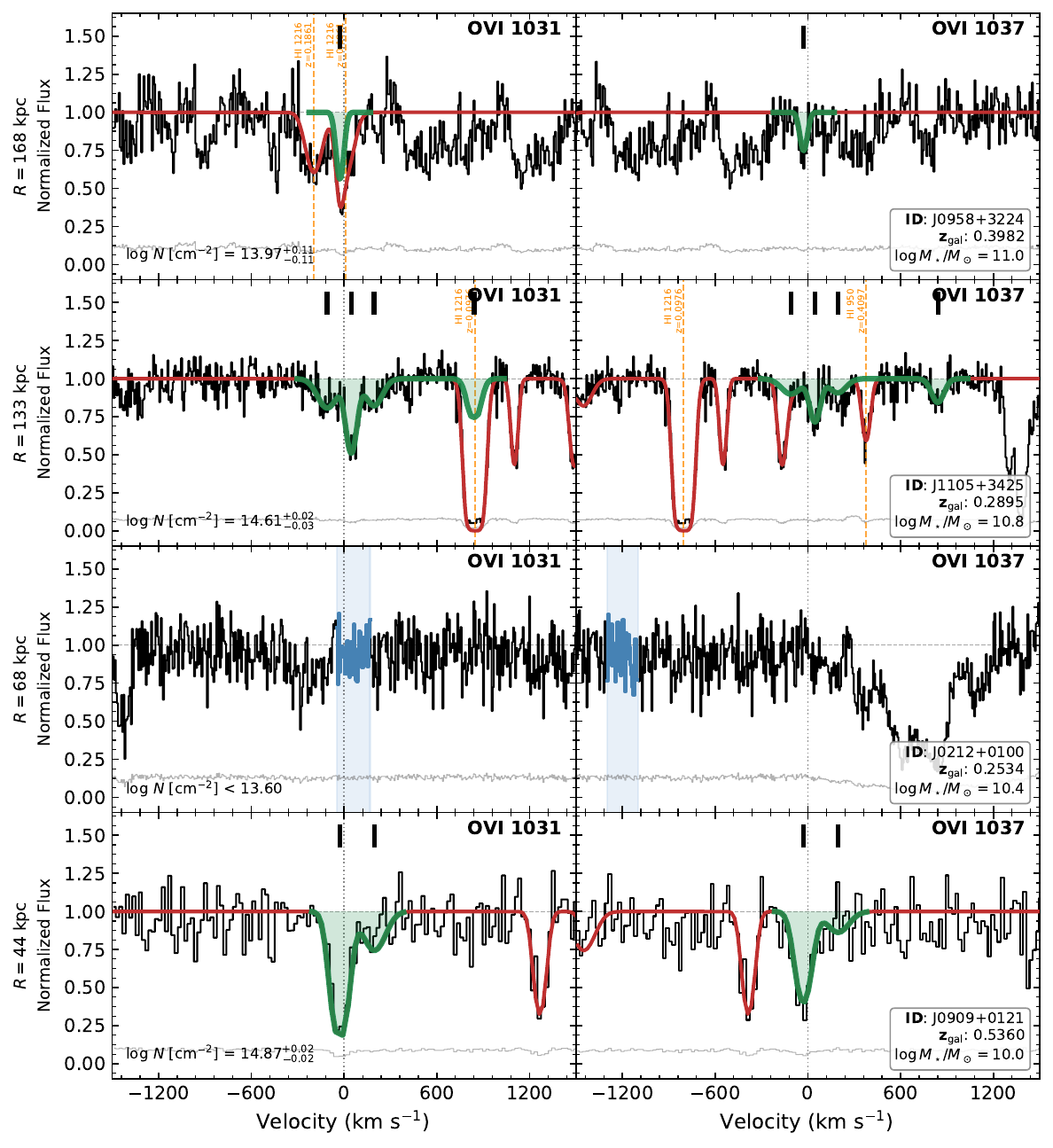}
    \caption{High resolution spectra of GOLIATH galaxies. Gratings used include G130M, G160M (some archival data includes G185M, and G225M). Detections have the complete Voigt profile fit plotted in red; \OVI\ components are plotted in green. Non-detections have the 200\kms\ window used for the $3\sigma$ upper limit calculation highlighted. Black tick marks represent the O VI kinematic components. Yellow dotted lines are blended components.}
    \label{fig:ovi_highres}
\end{figure}
\clearpage

% --- Low resolution figure (3 pages) ---
\begin{figure}[p]
    \centering
    \includegraphics[page=1, width=\linewidth]{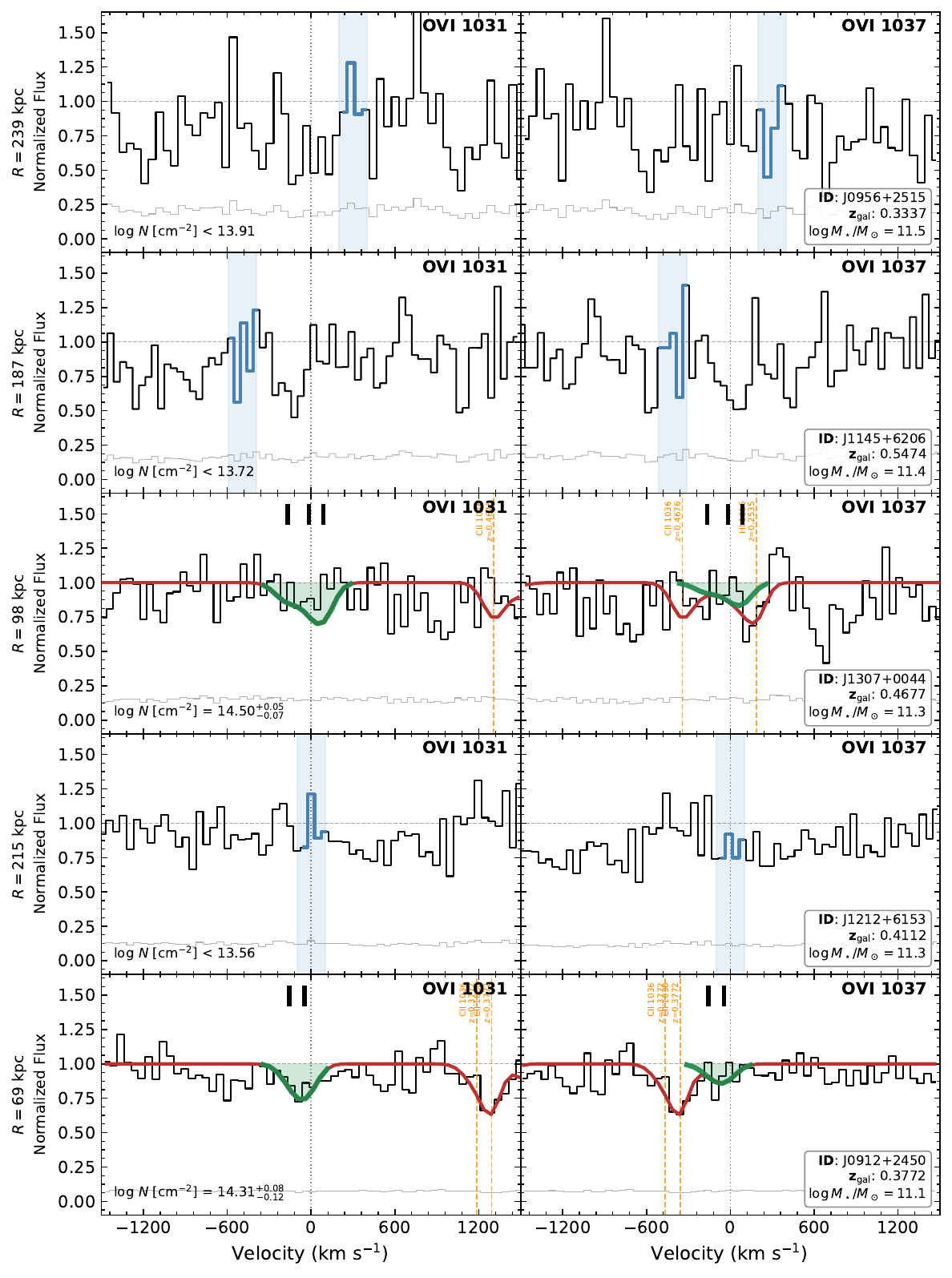}
\end{figure}
\clearpage
\begin{figure}[p]
    \ContinuedFloat
    \centering
    \includegraphics[page=2, width=\linewidth]{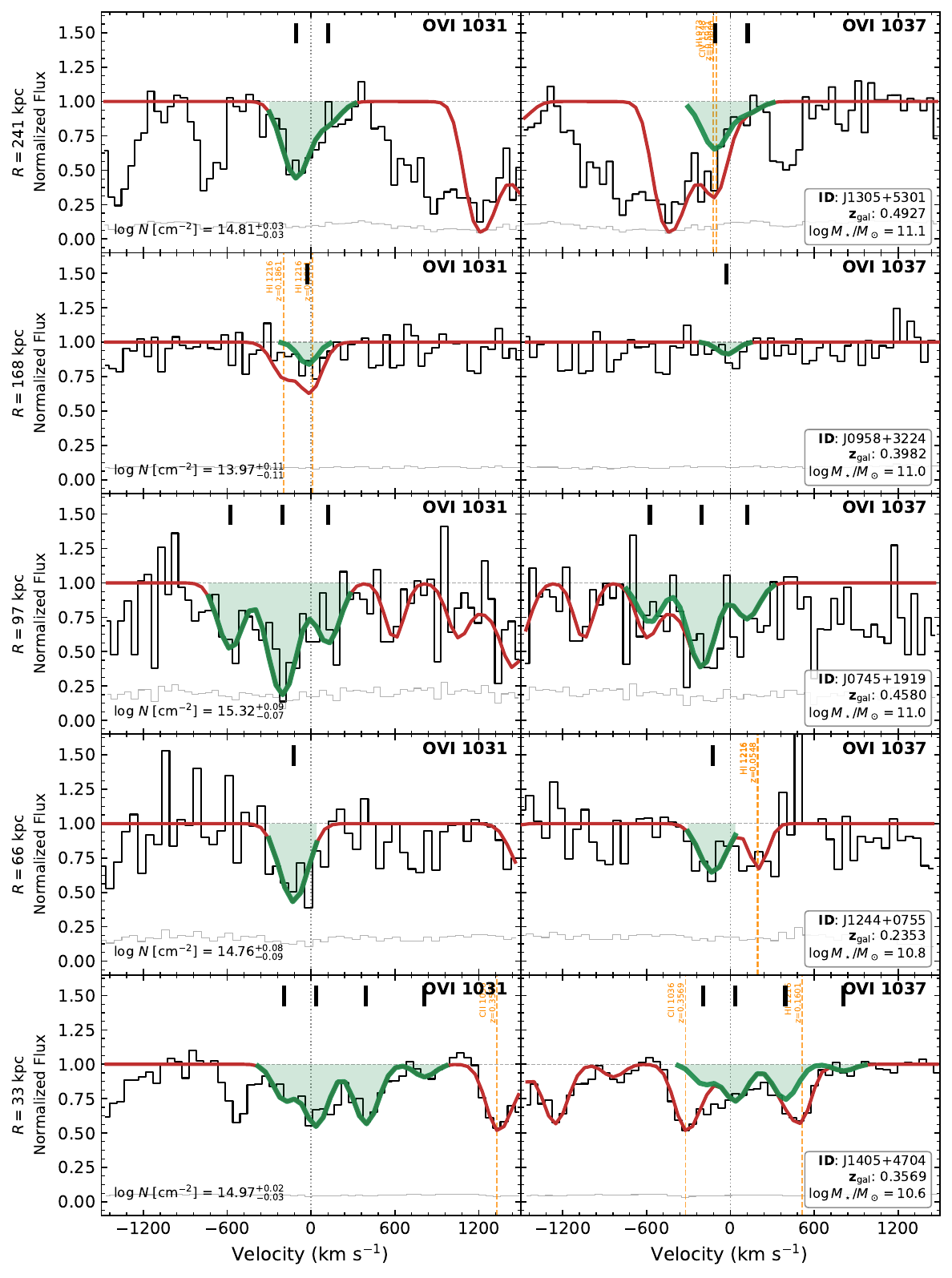}
\end{figure}
\clearpage
\begin{figure}[p]
    \ContinuedFloat
    \centering
    \includegraphics[page=3, width=\linewidth]{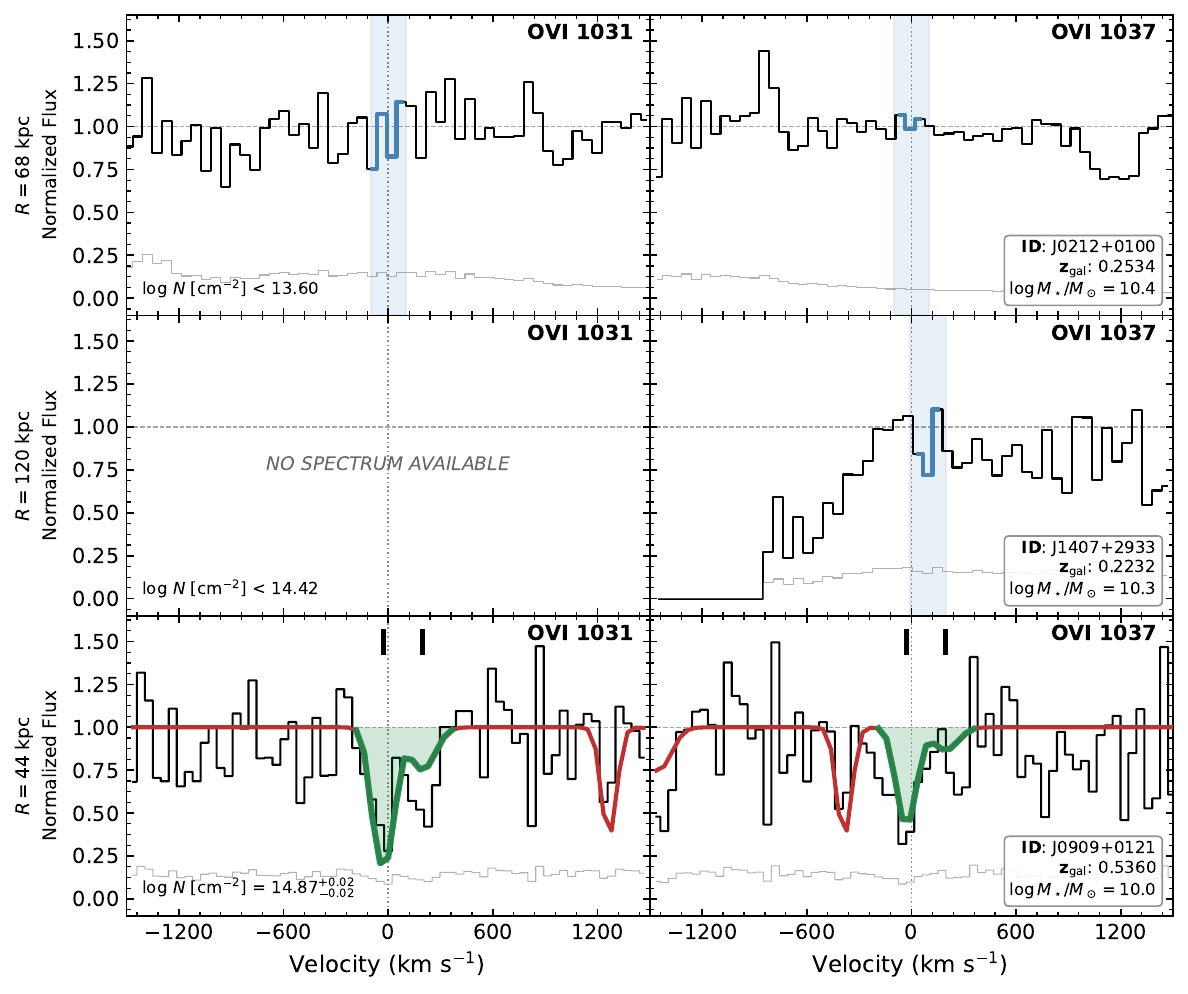}
    \caption{Low resolution spectra of GOLIATH galaxies. All observations use grating g140l. Detections have the complete Voigt profile fit plotted in red; green shows the \OVI\ components. Non-detections have the 200km/s window used for the $3\sigma$ upper limit calculation highlighted. Black tick marks represent the \OVI\ kinematic components. Yellow dotted lines are blended components.}
    \label{fig:ovi_lowres}
\end{figure}

\clearpage
\section{\OVI\ Component Breakdown}

\setlength{\tabcolsep}{6pt}
\startlongtable
\begin{deluxetable*}{llccccc}
\tablecaption{OVI Voigt Profile Parameters -- GOLIATH Survey\label{tab:ovi_voigt_bms}}
\tablehead{
\colhead{QSO ID} & \colhead{$z_\mathrm{gal}$} & \colhead{$\log\,(M_\star/M_\odot)$} &
  \colhead{\#$_{\rm comp}$} & \colhead{$\log N_{\rm O\,VI}$} & \colhead{$b$} & \colhead{$v_c$} \\
\colhead{} & \colhead{} & \colhead{} & \colhead{} & \colhead{(cm$^{-2}$)} & \colhead{(km\,s$^{-1}$)} & \colhead{(km\,s$^{-1}$)}
}
\startdata
% Detections
  J1319\text{+}2728 & $0.6704$ & $11.92$ & 5 & $14.01 \pm 0.05$ & \nodata & \nodata \\
  \hline
  & & & 1 & $14.31 \pm 0.03$ & $41.6^{+4.8}_{-4.0}$ & $-1764.9^{+1.9}_{-1.9}$ \\
  & & & 2 & $14.12^{+0.05}_{-0.06}$ & $82.1^{+2.2}_{-3.7}$ & $-1618.6^{+11.5}_{-9.5}$ \\
  & & & 3 & $13.33^{+0.16}_{-0.20}$ & $39.2^{+28.7}_{-18.9}$ & $-221.2^{+13.5}_{-11.2}$ \\
  & & & 4 & $13.66^{+0.06}_{-0.07}$ & $22.3^{+4.6}_{-4.6}$ & $151.9^{+3.0}_{-2.7}$ \\
  & & & 5 & $13.55^{+0.07}_{-0.08}$ & $14.9^{+5.1}_{-4.9}$ & $248.2^{+2.4}_{-2.2}$ \\
  \hline\hline
  J1307\text{+}0044 & $0.4677$ & $11.36$ & 3 & $14.50^{+0.06}_{-0.06}$ & \nodata & \nodata \\
  \hline
  & & & 1 & $13.95^{+0.11}_{-0.13}$ & $83.6^{+11.6}_{-17.4}$ & $-167.1^{+22.3}_{-19.6}$ \\
  & & & 2 & $13.84^{+0.14}_{-0.20}$ & $44.4^{+17.5}_{-17.0}$ & $-16.1^{+9.8}_{-11.2}$ \\
  & & & 3 & $14.20^{+0.06}_{-0.07}$ & $55.0^{+3.6}_{-6.6}$ & $87.8^{+7.2}_{-8.7}$ \\
  \hline\hline
  J1208\text{+}4540 & $0.9274$ & $11.23$ & 8 & $15.19^{+0.08}_{-0.09}$ & \nodata & \nodata \\
  \hline
  & & & 1 & $14.32^{+0.33}_{-0.71}$ & $46.4^{+23.0}_{-25.7}$ & $-381.6^{+23.3}_{-12.8}$ \\
  & & & 2 & $14.58^{+0.18}_{-0.40}$ & $67.4^{+9.8}_{-19.7}$ & $-322.8^{+25.8}_{-19.8}$ \\
  & & & 3 & $13.28^{+0.55}_{-0.64}$ & $25.4^{+22.4}_{-15.3}$ & $-215.9^{+31.0}_{-21.5}$ \\
  & & & 4 & $14.17^{+0.07}_{-0.16}$ & $50.2^{+6.5}_{-12.2}$ & $-157.9^{+10.9}_{-8.8}$ \\
  & & & 5 & $13.92^{+0.24}_{-0.35}$ & $15.4^{+13.6}_{-10.2}$ & $-22.8^{+6.9}_{-5.2}$ \\
  & & & 6 & $14.77 \pm 0.04$ & $78.5^{+7.5}_{-7.0}$ & $40.4^{+6.3}_{-6.0}$ \\
  & & & 7 & $13.81^{+0.18}_{-0.25}$ & $60.3^{+23.2}_{-25.6}$ & $176.5^{+17.3}_{-19.7}$ \\
  & & & 8 & $13.65^{+0.16}_{-0.24}$ & $62.0^{+6.2}_{-15.6}$ & $838.4^{+25.6}_{-62.5}$ \\
  \hline\hline
  J0912\text{+}2450 & $0.3772$ & $11.17$ & 2 & $14.31^{+0.08}_{-0.10}$ & \nodata & \nodata \\
  \hline
  & & & 1 & $13.70^{+0.20}_{-0.39}$ & $62.8^{+24.4}_{-23.2}$ & $-156.1^{+15.8}_{-24.0}$ \\
  & & & 2 & $14.19^{+0.08}_{-0.09}$ & $75.2^{+11.1}_{-16.9}$ & $-46.3^{+10.8}_{-13.5}$ \\
  \hline\hline
  J1342\text{-}0053 & $0.2271$ & $11.15$ & 3 & $14.54 \pm 0.03$ & \nodata & \nodata \\
  \hline
  & & & 1 & $14.10^{+0.05}_{-0.06}$ & $99.5^{+4.0}_{-7.3}$ & $-65.1^{+7.9}_{-3.6}$ \\
  & & & 2 & $14.24^{+0.04}_{-0.05}$ & $49.4^{+4.6}_{-4.9}$ & $36.5^{+3.1}_{-3.4}$ \\
  & & & 3 & $13.68^{+0.09}_{-0.09}$ & $28.6^{+10.0}_{-6.6}$ & $140.6^{+4.6}_{-6.9}$ \\
  \hline\hline
  J1305\text{+}5301 & $0.4927$ & $11.09$ & 2 & $14.81 \pm 0.03$ & \nodata & \nodata \\
  \hline
  & & & 1 & $14.74 \pm 0.03$ & $88.5^{+5.8}_{-6.0}$ & $-108.6^{+4.6}_{-5.0}$ \\
  & & & 2 & $14.00^{+0.07}_{-0.08}$ & $77.5^{+14.9}_{-15.4}$ & $124.1^{+4.2}_{-7.6}$ \\
  \hline\hline
  J0958\text{+}3224 & $0.3982$ & $11.05$ & 1 & $13.97 \pm 0.11$ & \nodata & \nodata \\
  \hline
  & & & 1 & $13.97 \pm 0.11$ & $31.3^{+9.3}_{-6.5}$ & $-27.0^{+3.8}_{-3.5}$ \\
  \hline\hline
  J0745\text{+}1919 & $0.4580$ & $10.99$ & 3 & $15.32^{+0.10}_{-0.07}$ & \nodata & \nodata \\
  \hline
  & & & 1 & $14.62 \pm 0.13$ & $79.5^{+15.3}_{-27.7}$ & $-575.8^{+32.3}_{-34.5}$ \\
  & & & 2 & $15.11^{+0.14}_{-0.09}$ & $87.2^{+9.5}_{-17.8}$ & $-205.6^{+18.0}_{-16.0}$ \\
  & & & 3 & $14.56^{+0.14}_{-0.16}$ & $75.0^{+17.5}_{-28.4}$ & $119.9^{+38.4}_{-35.1}$ \\
  \hline\hline
  J1244\text{+}0755 & $0.2353$ & $10.88$ & 1 & $14.76^{+0.08}_{-0.09}$ & \nodata & \nodata \\
  \hline
  & & & 1 & $14.76^{+0.08}_{-0.09}$ & $92.7^{+5.5}_{-11.3}$ & $-124.9^{+23.9}_{-23.7}$ \\
  \hline\hline
  J1105\text{+}3425 & $0.2895$ & $10.79$ & 4 & $14.60 \pm 0.03$ & \nodata & \nodata \\
  \hline
  & & & 1 & $13.95^{+0.04}_{-0.05}$ & $84.5^{+3.9}_{-7.8}$ & $-109.1^{+8.4}_{-9.9}$ \\
  & & & 2 & $14.18 \pm 0.03$ & $45.2^{+4.2}_{-2.9}$ & $46.4^{+2.2}_{-2.3}$ \\
  & & & 3 & $13.86 \pm 0.06$ & $74.1^{+10.5}_{-10.3}$ & $193.1^{+9.0}_{-8.2}$ \\
  & & & 4 & $13.96 \pm 0.08$ & $54.6^{+13.5}_{-10.0}$ & $840.6^{+8.5}_{-7.2}$ \\
  \hline\hline
  J1405\text{+}4704 & $0.3569$ & $10.64$ & 4 & $14.97 \pm 0.02$ & \nodata & \nodata \\
  \hline
  & & & 1 & $14.22 \pm 0.08$ & $57.5^{+18.4}_{-26.1}$ & $-193.7^{+2.6}_{-8.3}$ \\
  & & & 2 & $14.57 \pm 0.03$ & $78.3^{+1.3}_{-2.7}$ & $36.8^{+2.4}_{-4.9}$ \\
  & & & 3 & $14.55 \pm 0.03$ & $77.6^{+1.7}_{-4.4}$ & $393.1^{+8.2}_{-7.6}$ \\
  & & & 4 & $13.70^{+0.16}_{-0.24}$ & $49.1^{+22.0}_{-28.6}$ & $812.0^{+24.2}_{-32.2}$ \\
  \hline\hline
  J0909\text{+}0121 & $0.5360$ & $10.06$ & 2 & $14.87 \pm 0.02$ & \nodata & \nodata \\
  \hline
  & & & 1 & $14.78 \pm 0.02$ & $67.5^{+4.6}_{-4.4}$ & $-29.0^{+2.9}_{-2.8}$ \\
  & & & 2 & $14.14^{+0.05}_{-0.06}$ & $95.5^{+3.4}_{-6.8}$ & $195.8^{+15.9}_{-16.5}$ \\
  \hline\hline
% Non-detections
  J0956\text{+}2515 & $0.3337$ & $11.54$ & \nodata & $<13.88$ & \nodata & \nodata \\
  \hline\hline
  J1553\text{+}3548 & $0.4737$ & $11.49$ & \nodata & $<13.70$ & \nodata & \nodata \\
  \hline\hline
  J1145\text{+}6206 & $0.5474$ & $11.44$ & \nodata & $<14.04$ & \nodata & \nodata \\
  \hline\hline
  J1212\text{+}6153 & $0.4112$ & $11.31$ & \nodata & $<13.56$ & \nodata & \nodata \\
  \hline\hline
  J0212\text{+}0100 & $0.2534$ & $10.48$ & \nodata & $<13.60$ & \nodata & \nodata \\
  \hline\hline
  J1407\text{+}2933 & $0.2232$ & $10.30$ & \nodata & $<14.28$ & \nodata & \nodata \\
\enddata
\tablecomments{%
  Integrated $\log N$ is computed by summing component column densities in
  linear space; errors are propagated in quadrature from the 16th/84th
  percentile bounds of the MCMC posterior.
  Per-component broadening parameter ($b$) and velocity centroids ($v_c$) are the 50th-percentile (median) fit values
  with 16th/84th percentile errors.
  Sightline J1319$+$2728 only uses components within -300 and 200 km/s for the integrated column density as discussed in section \ref{sec:Results}.
  Upper limits denote $3\sigma$ equivalent-width limits converted to
  column density in the optically thin approximation.
  \nodata\ -- parameter not applicable for this row type.
}
\end{deluxetable*}

\end{document}